\pdfoutput=1

\documentclass{emulateapj}
\newcommand{\figurestar}{figure*}
\newcommand{\deluxetablestar}{deluxetable*} % for emulateapj
\usepackage{epstopdf}
\usepackage{amsmath}

\newcommand{\mc}{\multicolumn{2}{c}}
\newcommand{\mcf}{\multicolumn{4}{c}}

\newcommand{\Ha}{\ensuremath{\mathrm{H} \alpha}}

\newcommand{\Dxmin}{\ensuremath{\Delta x_{\mathrm{min}}}}
\newcommand{\Dxmax}{\ensuremath{\Delta x_{\mathrm{max}}}}

\newcommand{\cucm}{\ensuremath{\textrm{ cm}^{-3}}}

\newcommand{\kms}{\ensuremath{\textrm{ km s}^{-1}}}
\newcommand{\kpc}{\ensuremath{\, \mathrm{kpc}}}
\newcommand{\pc}{\ensuremath{\textrm{ pc}}}
\newcommand{\K}{\ensuremath{\textrm{ K}}}
\newcommand{\uG}{\ensuremath{\, \mu \mathrm{G}}}
\newcommand{\erg}{\ensuremath{\, \mathrm{erg}}}

\newcommand{\Myr}{\ensuremath{\textrm{ Myr}}}
\newcommand{\yr}{\ensuremath{\textrm{ yr}}}

\newcommand{\Bxini}{\ensuremath{B_{x,\mathrm{ini}}}}

\newcommand{\bxz}{{\tt bx0}}
\newcommand{\bxzhr}{{\tt bx0hr}}
\newcommand{\bxt}{{\tt bx10}}
\newcommand{\bxf}{{\tt bx50}}
\newcommand{\bxfhr}{{\tt bx50hr}}
\newcommand{\bxo}{{\tt bx100}}

\renewcommand{\vec}{\mathbf}

\newcommand{\pth}{\ensuremath{P_\mathrm{th}}}
\newcommand{\dd}[2]{\ensuremath{\frac{\partial #1}{\partial #2}}}

\newcommand{\Pturb}[1]{\ensuremath{{P_{\mathrm{turb},#1}}}}

\begin{document}

%\date{\today} % for aastex, not emulateapj

\submitted{ApJ, in press}

\title{Vertical structure of a supernova-driven turbulent, magnetized ISM}
\author{Alex S. Hill\altaffilmark{1,7}, M. Ryan Joung\altaffilmark{2,3}, Mordecai-Mark Mac Low\altaffilmark{3}, Robert A. Benjamin\altaffilmark{4}, L. Matthew Haffner\altaffilmark{1}, Christian Klingenberg\altaffilmark{5}, Knut Waagan\altaffilmark{6}}
\altaffiltext{1}{Department of Astronomy, University of Wisconsin-Madison, Madison, WI, USA; alex.hill@csiro.au}
\altaffiltext{2}{Department of Astronomy, Columbia University, New York, NY, USA}
\altaffiltext{3}{Department of Astrophysics, American Museum of Natural History, New York, NY, USA}
\altaffiltext{4}{Department of Physics, University of Wisconsin-Whitewater, Whitewater, WI, USA}
\altaffiltext{5}{Department of Mathematics, W\"{u}rzburg University, Emil Fischer Strasse 30, W\"{u}rzburg, Germany}
%\altaffiltext{6}{The University of Maryland, CSCAMM, College Park MD, USA}
\altaffiltext{6}{%Present address:
Department of Applied Mathematics, University of Washington, Seattle, WA, USA}
\altaffiltext{7}{Present address: CSIRO Astronomy \& Space Science, Marsfield, NSW, Australia}

\begin{abstract}
Stellar feedback drives the circulation of matter from the disk to the halo of galaxies. We perform three-dimensional magnetohydrodynamic simulations of a vertical column of the interstellar medium with initial conditions typical of the solar circle in which supernovae drive turbulence and determine the vertical stratification of the medium. The simulations were run using a stable, positivity-preserving scheme for ideal MHD implemented in the FLASH code. We find that the majority ($\approx 90 \%$) of the mass is contained in thermally-stable temperature regimes of cold molecular and atomic gas at $T < 200 \K$ or warm atomic and ionized gas at $5000 \K < T < 10^{4.2} \K$, with strong peaks in probability distribution functions of temperature in both the cold and warm regimes. The $200 - 10^{4.2} \K$ gas fills $50-60 \%$ of the volume near the plane, with hotter gas associated with supernova remnants ($30-40 \%$) and cold clouds ($< 10 \%$) embedded within. At $|z| \sim 1-2 \kpc$, transition-temperature ($10^5 \K$) gas accounts for most of the mass and volume, while hot gas dominates at $|z| > 3 \kpc$. The magnetic field in our models has no significant impact on the scale heights of gas in each temperature regime; the magnetic tension force is approximately equal to and opposite the magnetic pressure, so the addition of the field does not significantly affect the vertical support of the gas. The addition of a magnetic field does reduce the fraction of gas in the cold ($< 200 \K$) regime with a corresponding increase in the fraction of warm ($\sim 10^4 \K$) gas. However, our models lack rotational shear and thus have no large-scale dynamo, which reduces the role of the field in the models compared to reality. The supernovae drive oscillations in the vertical distribution of halo gas, with the period of the oscillations ranging from $\approx 30 \Myr$ in the $T < 200 \K$ gas to $\sim 100 \Myr$ in the $10^6 \K$ gas, in line with predictions by \citeauthor{wc01}.
\end{abstract}

\maketitle

\section{Introduction}

Supernovae regulate the structure and dynamics of the multi-phase interstellar medium \citep[ISM;][]{cs74,mo77}. They produce hot ($\sim 10^6 \K$) gas that occupies a large, though uncertain, fraction of the volume in the disk and halo. Supernovae drive circulation of gas from the disk to heights of several kiloparsecs in a so-called ``Galactic fountain'' \citep{sf76,b80}. They also drive turbulence \citep{nf96,mk04}, and a supernova-driven turbulent dynamo can amplify a very small magnetic field to a significant fraction of equipartition \citep{bkm04}. Turbulent and magnetic energy densities dominate over the thermal energy in many interstellar environments, particularly in cold atomic and molecular gas \citep{m90,ht05}, and provide, along with high-energy particles, a crucial component of the vertical pressure that supports the ISM in hydrostatic equilibrium \citep{bc90}. Thus, supernovae likely provide the momentum that is responsible for the majority of the pressure support for the ISM.

A complete understanding of the interplay between components of the ISM requires modeling of supernova-driven magnetized turbulence in a multi-phase medium. To connect the wide range of temperatures ($10 - 10^7 \K$), densities ($10^{-4} - 10^3 \cucm$), and size scales (parsecs to kiloparsecs) found in the diffuse ISM, simulations must have enough spatial resolution near the plane to resolve individual supernova explosions early in their evolution so that unphysical radiative cooling does not remove the energy injected by the supernovae. The simulations also must extend to large enough heights that the mass flux out the top of the simulation box is reasonably small, as well as keep track of heating and cooling of the gas. Variable mesh techniques such as nested grid or adaptive mesh refinement (AMR) are essential to resolve the small-scale structure near the plane while covering a sufficient vertical volume of the Galaxy with available computational resources.

Hydrodynamical and magnetohydrodynamical (MHD) studies of this sort in two \citep{cp85,rb95,rbk96,ko09,ko09a} and three (\citealt{kbs99,ab04,ab05}; \citealt[hereafter JM06]{jm06}; \citealt{gez08,gze08}; \citealt[hereafter JMB09]{jmb09}) dimensions (3D) have had considerable success in producing the observed cold, warm neutral and ionized \citep{whj10}, and hot gas with reasonable distributions of mass between these components of the ISM near the plane of the Galaxy. However, comparisons to observations suggest three inconsistencies.

First, the observed distribution of emission measures from the warm ionized medium (WIM) is narrower than the models, suggesting that the real ISM is more compressible than the simulations \citep{hbk08,whj10}. Second, the simulated WIM disk is thinner than observed \citep{whj10}. Third, the observed X-ray temperature of halo gas is cooler than predicted by the models \citep{hsk10}. Because we expect magnetic fields reduce the compressibility of a fluid and provide vertical support, the addition of magnetic fields may help to address all of these inconsistencies.

With these motivations, we extend the simulations developed by JMB09 to use MHD in this paper. Our primary aim here is to explore the effect of magnetic fields on the vertical structure of the multiphase ISM. In future work, we will conduct Monte Carlo photoionization simulations \citep[following][]{whj10} of the models presented here and compare the results to \Ha, Faraday rotation, and \ion{H}{1} observations. In Section~\ref{sec:simulations}, we discuss our simulations. In Section~\ref{sec:results}, we describe the results of the simulations, including the evolution over time (Section~\ref{sec:time_evol}), the phases of the ISM (Section~\ref{sec:gas_phases}), the vertical structure (Section~\ref{sec:vertical_structure}), the magnetic field energy and pressure (Section~\ref{sec:b}), and the thermal and turbulent pressure (Section~\ref{sec:pressure}). We discuss our results in Section~\ref{sec:sn_mhd_discussion}, focusing on the phases of the ISM in Sections~\ref{sec:phases_heating}--\ref{sec:pressure_discussion}. We discuss the vertical stratification in Section~\ref{sec:stratification}, with a comment on the substantial relevance of the \citet{wc01} models to the evolution of the halo in our results in Section~\ref{sec:oscillation}. We finish with discussion of the generation of the magnetic field (Section~\ref{sec:dynamo}), the effects of numerical resolution (Section~\ref{sec:numerical_res}), and some large-scale effects which are not included in our model (Section~\ref{sec:missing_largescale}). We summarize our main conclusions in Section~\ref{sec:sn_mhd_conclusions}.

\section{Simulations} \label{sec:simulations}

\subsection{Model} 
Our model is an extension of the hydrodynamical model developed by JM06 and JMB09. We use FLASH v2.5, an Eulerian astrophysical code with AMR \citep{for00} in 3D. We employ the positivity-preserving HLL3R MHD Riemann solver based scheme developed and implemented in FLASH by \citet{ksw07,w09,kw10}; and \citet{wfk11}. The scheme is stable and preserves positivity even for the Mach numbers approaching 100 and very low values of plasma $\beta$ ($< 0.1$) found in this problem. The code solves the ideal MHD equations, written here in Gaussian cgs units:
\begin{align}
\dd{\rho}{t} + \nabla \cdot (\rho \vec{v}) = {} & 0 \label{eq:masscons} \\
  \begin{split}
	\dd{\rho \vec{v}}{t} + \vec{\nabla} \cdot \left[\rho \vec{v} \vec{v} + \left(\pth + \frac{|\vec{B}|^2}{8 \pi} \right) \vec{I} - \right. \\
	\left.
\frac{\vec{B} \vec{B}}{4 \pi} \right] = {} & \rho g(z) \vec{\hat{z}} \label{eq:momentumcons} 
  \end{split} 
\end{align}
\begin{align}
\begin{split}
	\dd{E}{t} + \vec{\nabla} \cdot \left[ \left( E + \frac{|\vec{B}|^2}{8 \pi} + \pth \right) \vec{v} - \right. & \\
	\left.
\frac{(\vec{B} \cdot \vec{v}) \vec{B}}{4 \pi} \right] = {} & n \Gamma - n^2 \Lambda + S \label{eq:energycons}
\end{split} \\
\dd{\vec{B}}{t} - \vec{\nabla} \times \left( \vec{v} \times \vec{B} \right) = {} & 0 \label{eq:magneticfluxcons}.
\end{align}
Equations (\ref{eq:masscons}) -- (\ref{eq:magneticfluxcons}) represent conservation of mass density $\rho$, momentum $\rho \vec{v}$, energy, and magnetic flux. Here $\mathbf{I}$ is the $3 \times 3$ identity matrix, and the notation $\vec{B}\vec{B}$ denotes the outer product of $\vec{B}$ with itself. The constraint $\vec{\nabla} \cdot \vec{B} = 0$ applies. The thermal pressure is $\pth$ (with a $\gamma = 5/3$ equation of state), the energy density
\begin{equation} \label{eq:energy}
E = \frac{\rho |\vec{v}|^2}{2} + \frac{\pth}{\gamma - 1} + \frac{|\vec{B}|^2}{8 \pi},
\end{equation}
and the magnetic field $\vec{B}$. The diffuse heating rate $n \Gamma(z,T)$, the cooling rate $n^2 \Lambda(T)$, and the impulsive heating due to supernovae and stellar winds in superbubbles $S(\vec{x}, t)$ are as specified in JM06 and summarized below. We discuss the gravitational acceleration $g(z)$ below; $\vec{\hat{z}}$ is a unit vector. Although we do not include explicit Ohmic resistivity in the model, numerical resistivity is present.

\begin{deluxetable}{r@{.}l @{$\, < |z| < \,$} r@{.}l r@{.}l r@{.}l}
\tabletypesize{\footnotesize}
\tablecolumns{8}
\tablewidth{0pt}
\tablecaption{Refinement zones}
\tablehead{\mcf{$|z|$ (kpc)} & \mcf{$\Delta x$ (pc)} \\
\mcf{} & \mc{high res} & \mc{medium res} }
\startdata
0&		& 0&05		& 1&95	& 3&91 	\\
0&05	& 0&3		& 3&91	& $\le7$&81 	\\
0&3		& 1&		& 7&81	& 15&6 	\\
1&		& 10&		& 15&6	& 15&6	\\
10&		& 20&		& 31&2	& 31&2 
\enddata
\label{tbl:amr}
\end{deluxetable}

The simulation consists of an elongated box $1 \times 1 \times 40 \kpc^3$ in size. The midplane is at the center, with the box extending to $z = \pm 20 \kpc$. We employ zero-gradient boundary conditions (as described in Section~\ref{sec:box_size} below) at the top and bottom of the box and periodic boundary conditions through the sides. The initial conditions consist of gas in approximate hydrostatic equilibrium. The temperature at the midplane is $1.15 \times 10^4 \K$, while the initial temperature above $|z| \approx 1 \kpc$ is $1.15 \times 10^6 \K$. In this paper, we report hydrogen number densities $n_\mathrm{H} \equiv n \equiv \rho / \bar{m}$ adopting a mean mass per hydrogen atom $\bar{m} = 2.36 \times 10^{-24} \textrm{ g}$, appropriate for a solar metallicity gas. The adaptive mesh is finest near the midplane and becomes coarser with height, as shown in Table~\ref{tbl:amr}.

Parameters describing our runs are listed in Table~\ref{tbl:runs}. Our naming convention consists of the initial, midplane, horizontal ($x$) component of the magnetic field in code units, with {\tt hr} added for runs with a resolution of $\approx 2 \pc$; runs without the {\tt hr} label are medium ($\approx 4 \pc$) resolution. We run the medium resolution setup for $400 \Myr$. For two models (\bxzhr\ and \bxfhr), we also switch to high resolution after $300 \Myr$. The medium resolution run establishes an improved initial condition over the hydrostatic equilibrium setup for the high resolution run at relatively low computational cost. For the majority of this paper, we focus on the results from the \bxzhr, \bxfhr, and \bxo\ models, which have initial midplane magnetic fields of $0$, $6.5$, and $13 \uG$, respectively. The \bxzhr\ and \bxfhr\ models ran from $300 \Myr$ to at least $340 \Myr$, while the \bxo\ model ran to $260 \Myr$. Runs were conducted on the Ranger cluster at the Texas Advanced Computing Center; $4 \pc$ resolution runs utilized $\sim 80-200$ service units per Myr of evolution, while $2 \pc$ resolution runs required $\sim 1500-2500$ service units per Myr.

\begin{deluxetable}{lr@{.}l c lr rr}
\tabletypesize{\footnotesize}
\tablecolumns{8}
\tablewidth{0pt}
\tablecaption{List of runs}
\tablehead{\colhead{Name} & \mc{$\Bxini$} & \colhead{$T_\mathrm{halo}$} & \colhead{\Dxmin} & \colhead{\Dxmax} & \colhead{$t_\mathrm{start}$} & \colhead{$t_\mathrm{end}$} \\
\colhead{} & \mc{(\uG)} & \colhead{$(1.15 \K)$} & \colhead{(pc)} & \colhead{(pc)} & \colhead{(Myr)} & \colhead{(Myr)} }
\startdata
\bxz	& 0&0		& $10^6$	& $3.91$	& $31.2$	& 0	& 400	\\
\bxzhr	& 0&0		& $10^6$	& $1.95$	& $31.2$	& 300	& 340	\\
\bxt	& 1&3		& $10^6$	& $3.91$	& $31.2$	& 0	& 270	\\
\bxf	& 6&5		& $10^6$	& $3.91$	& $31.2$	& 0	& 400	\\
\bxfhr	& 6&5		& $10^6$	& $1.95$	& $31.2$	& 300	& 370	\\
\bxo	& 13&		& $10^6$	& $3.91$	& $31.2$	& 0	& 260
\enddata
\tablecomments{The initial temperature at heights $|z| \gtrsim 1 \kpc$ is shown as $T_\mathrm{halo}$. }
\label{tbl:runs}
\end{deluxetable}

Supernova explosions are set off in the simulation box as described by JM06, adding $10^{51} \erg$ to a sphere encompassing $60 \, M_\odot$ centered on the supernova location and redistributing the mass within the sphere. No gas mass is added. This approach prevents excessive cooling in the timesteps immediately after the supernova. We use the Galactic supernova rate from JMB09, with Type Ia and core-collapse supernova rates of $6.58$ and $27.4 \Myr^{-1} \kpc^{-2}$, respectively. Three-fifths of the core-collapse supernovae are clustered spatially and temporally to simulate superbubbles. The number of supernovae in each superbubble is drawn from a distribution $dN_B \propto n_*^{-2} dn_*$ between upper and lower cutoffs $n_{\mathrm{sn,max}} = 40$ and $n_{\mathrm{sn,min}} = 7$, respectively. During the first $5 \Myr$ of each superbubble, additional energy is injected to simulate stellar winds (see JM06). We assign the Type Ia and core-collapse supernovae exponential vertical distributions with scale heights of $325$ and $90 \pc$, respectively. The locations of supernova explosions are random with these restrictions, not correlated with the gas density. The supernova rate does not change over time; this is equivalent to a constant star formation rate. The random times and locations of supernovae are chosen at initialization ($t=0$), so the supernovae occur in the same locations for all runs with a given configuration independent of resolution: the supernovae occur at the same times and locations in \bxz\ and \bxzhr\ models, but the supernovae occur at different times and locations in the \bxz\ and \bxf\ runs.

As in JM06 and JMB09, we use radiative cooling appropriate for an optically thin, solar metallicity plasma in collisional ionization equilibrium \citep[see Figure 1 of JM06]{dm72,sd93}. We apply a diffuse heating term representing photoelectric heating of dust grains \citep{bt94}, thought to be the dominant heating mechanism in the warm and cold neutral medium (WNM and CNM, respectively). The diffuse heating rate is  \citep{whm95}
\begin{equation} \label{eq:heating}
\Gamma(z,T) \equiv \left\{ \begin{array}{rl}
\epsilon G_0 e^{-|z| / h_\mathrm{pe}} \times 10^{-24} \textrm{ erg s}^{-1}, & T < 2 \times 10^4 \K \\
 0, & T > 2 \times 10^4 \K.
\end{array} \right.
\end{equation}
We employ a heating efficiency $\epsilon = 0.05$, a midplane FUV intensity $G_0=1.7$ times the interstellar value near the Sun \citep{h68}, and a FUV radiation field scale height $h_\mathrm{pe} = 300 \pc$.

We do not include any parameterized photoionization heating, as we discuss in Section~\ref{sec:missing}.

\subsubsection{Magnetic field}

The primary modification we make to the physics included in the JM06 and JMB09 models is the addition of magnetic fields. \citet{wfk11} present numerous astrophysical tests of the HLL3R MHD solver used in this work. We find that a Courant step of $0.2$ is required early in the simulations, as shocks are first established, to obtain a stable solution. After $\sim 10 \Myr$, we increase the Courant step to $0.4 - 0.8$ to improve performance and find that the resulting solutions are generally stable. Densities and temperatures always remain positive by construction; unstable solutions are typically alleviated by manually reducing the time step.

The magnetic field is established with an initially uniform, horizontal field with a constant plasma $\beta$, i.~e.\ $\Bxini(z) = \Bxini(z=0) \times \left[ n(z) / n(z=0) \right]^{1/2} \equiv \Bxini$. Although the random component of the magnetic field in the Solar neighborhood is thought to be $\sim 3$ times the strength of the uniform field \citep[e.~g.][]{f01}, we do not initialize any random component, allowing it to be generated by turbulence. The initial, midplane magnetic field strengths for each of our runs are listed in Table~\ref{tbl:runs}.

\subsubsection{Box size} \label{sec:box_size}

We find that the vertical extent of the box is the most important numerical factor in establishing a realistic temperature distribution in the halo. In particular, the Galactic fountain flow is not accurately captured in simulations presented by JMB09 that only extend to $z = \pm 5 \kpc$ \citep[cf.][]{ab04} and do not enforce outflow at the boundary. In these runs, hot gas from Type~Ia supernovae above $|z| \sim 500 \pc$ in the first few Myr of the simulation moves rapidly upward. Pressure and density from the top row of the box ($z = \pm 5 \kpc$) are copied to the ghost cells just outside the box. When the pressure inside later drops, after passage of the blast wave, a constant inflow of hot, high pressure gas is established, unphysically creating ram pressure that heats the gas at $|z| \gtrsim 1 \kpc$ to temperatures of $\sim 10^7 \K$. \citet{jbp11} discuss this numerical issue further. We extend our box to heights $|z| = 20 \kpc$ with very low ($32 \pc$) resolution at $|z| > 10 \kpc$. With this box size, we track enough of the mass and energy flux in the fountain flow so that little mass actually flows off the grid.

We employ a gas surface mass density of $13.2 \, M_{\odot} \pc^{-2}$, the approximate value at the solar circle \citep{fhp06}. The star formation rate inferred from the supernova rate we use does not lie on the Kennicutt-Schmidt law \citep[see also JMB09]{keh89}.

\subsubsection{Gravitational potential} \label{sec:potential}

\begin{figure}
\plotone{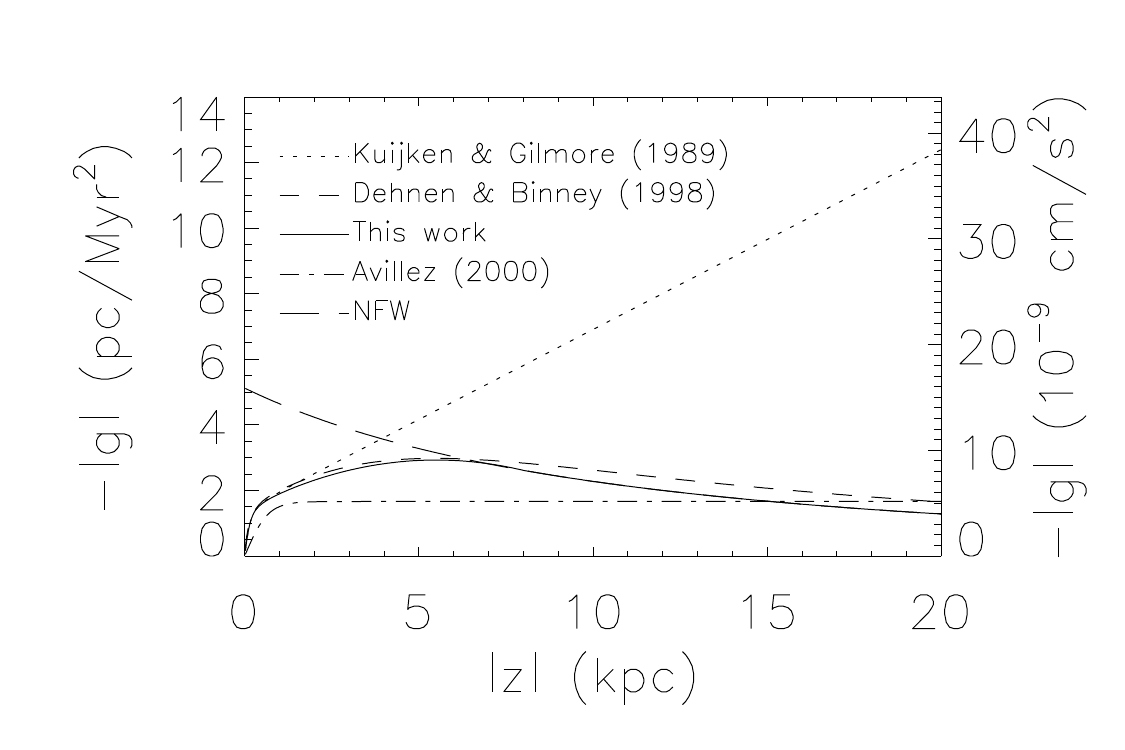}
\caption{Gravitational acceleration using potentials from \citet[][as used by JM06 and JMB09]{kg89}, \citet{db98} with a spheroidal dark matter halo, \citet{a00} (assuming the $\rho_*$ in their equation (2) should be $\rho_{*,0}$), and \citet[][NFW]{nfw96} (normalized to match equation~[\ref{eq:pot}] at $|z| = 8 \kpc$). The potential used in this work (described in Section~\ref{sec:potential}) is shown with a solid line.}
\label{fig:pot}
\end{figure}

We use a modified version of the \citet{kg89} gravitational potential, consisting of contributions from a stellar disk and a spherical dark halo:
\begin{equation} \label{eq:pot}
g(z) = -\frac{a_1 z}{\sqrt{z^2 + z_0^2}} - a_2 z + a_3 z |z|, |z| \le 8 \kpc.
\end{equation}
The constants $a_1 = 1.42 \times 10^{-3} \kpc \Myr^{-2}$, $a_2 = 5.49 \times 10^{-4} \Myr^{-2}$, and $z_0 = 0.18 \kpc$. In Figure~\ref{fig:pot}, we show the unmodified ($a_3 = 0$) \citet{kg89} gravitational acceleration used by JM06 and JMB09. We also show the acceleration due to a parameterized model of the mass distribution of the Milky Way, fitted to the available data by \citet{db98}, at the solar circle (Galactocentric radius $R=R_0=8 \kpc$). Because of the limited sample of observed stellar radial velocities, the shape of the dark matter halo is uncertain. However, the linear acceleration of the \citet{kg89} potential (dotted line in Fig.~\ref{fig:pot}) overestimates the observed \citep{db98} acceleration for a flat halo (short dashed line) at $|z| \gtrsim 4 \kpc$.

The shape of the dark halo of the Galaxy is unknown; claims of an oblate, spheroidal, and prolate halo have all been made \citep[e.\ g.][]{h04,jlm05,fbe06,lm10}. Therefore, we add the quadratic term in equation~(\ref{eq:pot}) to the acceleration derived from \citet{kg89} by setting $a_3 = 5 \times 10^{-5} \kpc^{-1} \Myr^{-2}$, yielding a value of $g(z)$ roughly consistent with the \citet{db98} constraints in the range $|z| < 8 \kpc$. Note that this functional form is unphysical at heights $|z| \gtrsim 8 \kpc$. At these heights, the gravitational potential is dominated by the dark matter halo. Above $|z| = 8 \kpc$, we use the acceleration of a \citet{nfw96} spherical dark matter profile with a scale length of $20 \kpc$ normalized to match equation~(\ref{eq:pot}) at $|z| = 8 \kpc$. Because of the large uncertainties in the Galactic potential at high $|z|$, we do not attempt a more detailed fit.

\subsection{Missing components} \label{sec:missing}

Our models are designed to incorporate the dominant physical processes at work in the diffuse ISM. As discussed by JM06, we do not include the effects of self-gravity, thermal conduction, differential rotation, or a radial variation in the gravity. We also do not include the effect of high-energy particles (cosmic rays), which provide vertical pressure comparable to the kinetic and magnetic pressure in the solar neighborhood \citep{bc90,f01}.
Stellar feedback is only incorporated through supernovae, the stellar wind during the first $5 \Myr$ of the lifetime of each superbubble, and the diffuse heating term. Smaller-scale, lower energy drivers of turbulence and structure formation such as expanding \ion{H}{2} regions \citep{m02,bhm09} are not included. We note that \citet{ko09,ko09a} include \ion{H}{2} regions but not supernovae in their global 2D simulations.

We do not track UV radiation, so neither photoelectric heating nor photoionization heating are accounted for properly. The diffuse heating term discussed above approximates photoelectric heating by the FUV radiation field in a time- and space-averaged sense. However, \citet{phm03} showed that the UV radiation field and photoelectric heating are likely highly time-variable in the solar neighborhood. We comment on the effects of this in Section~\ref{sec:phases_heating}. The absence of UV radiation in our models also means we cannot track photoionization, meaning that we can only distinguish between the WNM and WIM as a post-processing step \citep{whj10}. The thermal pressure for a given mass density in the WIM is twice that in the WNM due to the extra particle (electron) associated with each hydrogen atom. The impact of this on the dynamics may be small because the thermal pressure is generally significantly smaller than the turbulent pressure in $10^4 \K$ gas, as shown in Section~\ref{sec:pressure} below. However, we also cannot include photoionization heating, an important heating source in the WIM \citep{rht99}. Without photoionization heating, the difference between the thermal pressure in real warm ionized gas and that in our model may be more than the factor of two mentioned above due to the increased temperature. An increase in the diffuse heating rate (equation \ref{eq:heating}) or a second component with a larger scale height could mimic photoionization heating but is not included. We will discuss the effects of UV radiation on the heating rate further in future work.

\section{Results} \label{sec:results}

\subsection{Evolution in the halo} \label{sec:time_evol}

\begin{\figurestar}
\plotone{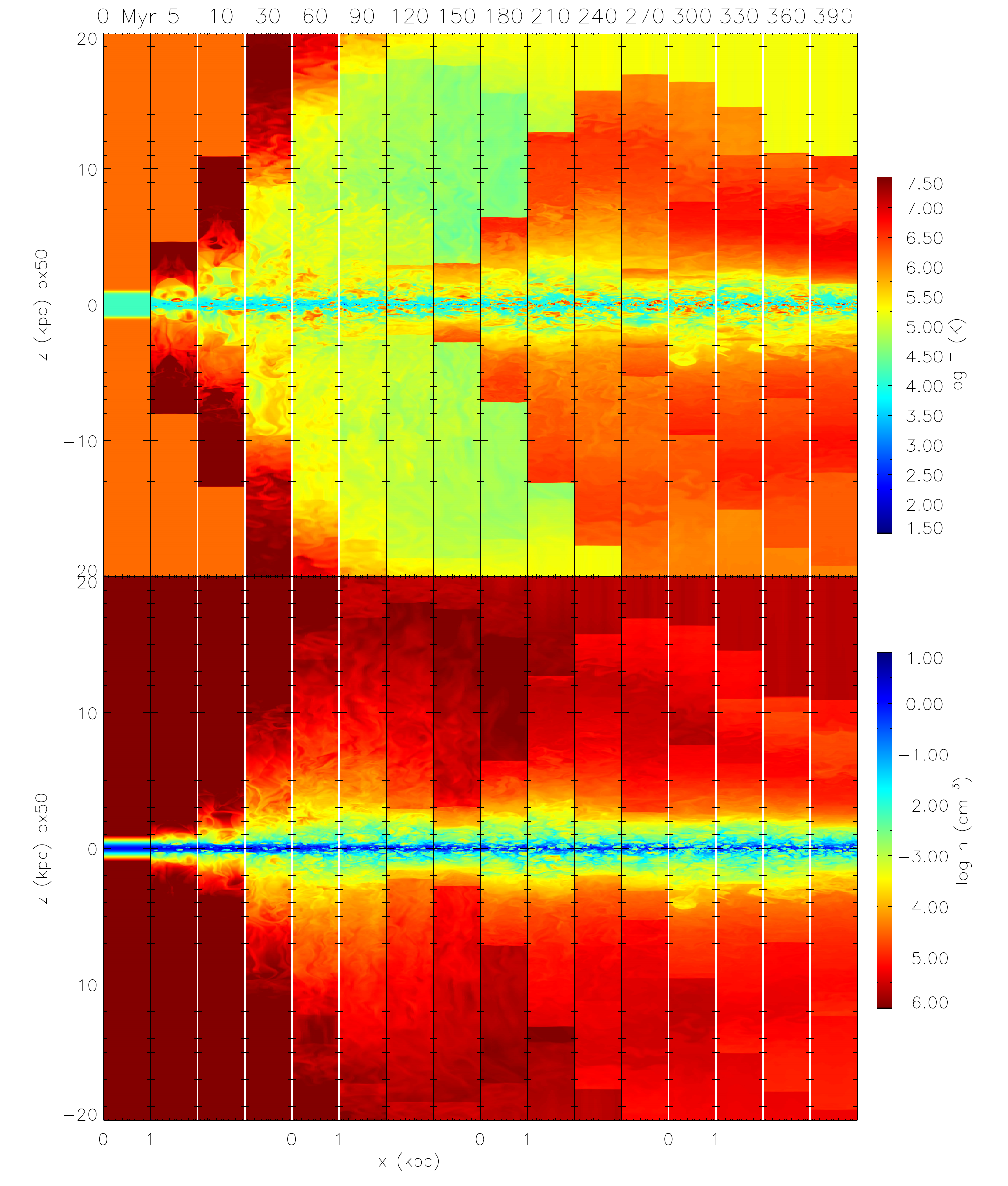}
\caption{Slices of temperature ({\em top}) and density ({\em bottom}) over time for the \bxf\ model. The time of each slice (in Myr) is listed above each pair of images. Note that the aspect ratio is not unity; the vertical scale of these images is contracted. The horizontal ($x$) axis values are labeled for four slices only to reduce clutter.}
\label{fig:images_evol}
\end{\figurestar}

The evolution of the vertical structure of the \bxf\ model of the ISM is shown through slices of temperature and density over time in Figure~\ref{fig:images_evol}. The first supernovae that explode at heights $|z| \approx 1 \kpc$ occur in a sufficiently low-density environment that they produce shocks that propagate upwards, heating the entire volume. Because of our chosen initial condition, the densities at heights $|z| \gg 1 \kpc$ are $\approx 10^{-6} \cucm$. This is much lower than typical halo densities of $\sim 10^{-4} \cucm$, which can be estimated from dispersion measures of pulsars in the Magellanic Clouds \citep[e.~g.][]{mfl06}. Therefore, shocks heat the gas more than would be expected with a more realistic initial condition. The initial shocks escape our box after $\approx 20 \Myr$.

The post-shock gas at very large heights ($|z| \gtrsim 10 \kpc$) cools quickly to temperatures of $\sim 10^{4.6} \K$, a local minimum in our cooling curve. At lower heights, denser gas is forced upwards. This gas is only heated to temperatures of $\sim 10^5 \K$ and reaches heights of $\lesssim 10 \kpc$, consistent with a supernova-driven Galactic fountain; we chose our box size to fully capture the cycling of this material.

The fountain material ejected early in our simulations returns to the plane after $\approx 100 \Myr$. At this point, a second pair of shocks propagates away from the plane (see Section~\ref{sec:stratification} and \citealt{wc01}).\footnote{We emphasize that our supernova rate is constant with time, so new star formation is not triggered by the falling fountain material.} After this second pair of shocks passes through the medium, the halo density is more consistent with observations, so reasonable initial conditions are now established. Low resolution simulations confirm that the impact of the initial conditions is negligible after this point. This second pair of shocks heats gas in the halo ($|z| \gtrsim 3 \kpc$) to temperatures of $\sim 10^6 \K$. A layer from $0.5 \kpc \lesssim |z| \lesssim 2 \kpc$ consists primarily (by volume) of thermally unstable, $10^5 \K$ gas, while warm ($\sim 10^4 \K$) gas dominates (by volume) closer to the plane. We discuss these filling fractions as a function of height below.

\subsection{Gaseous components of the ISM} \label{sec:gas_phases}

\begin{\figurestar}
\plotone{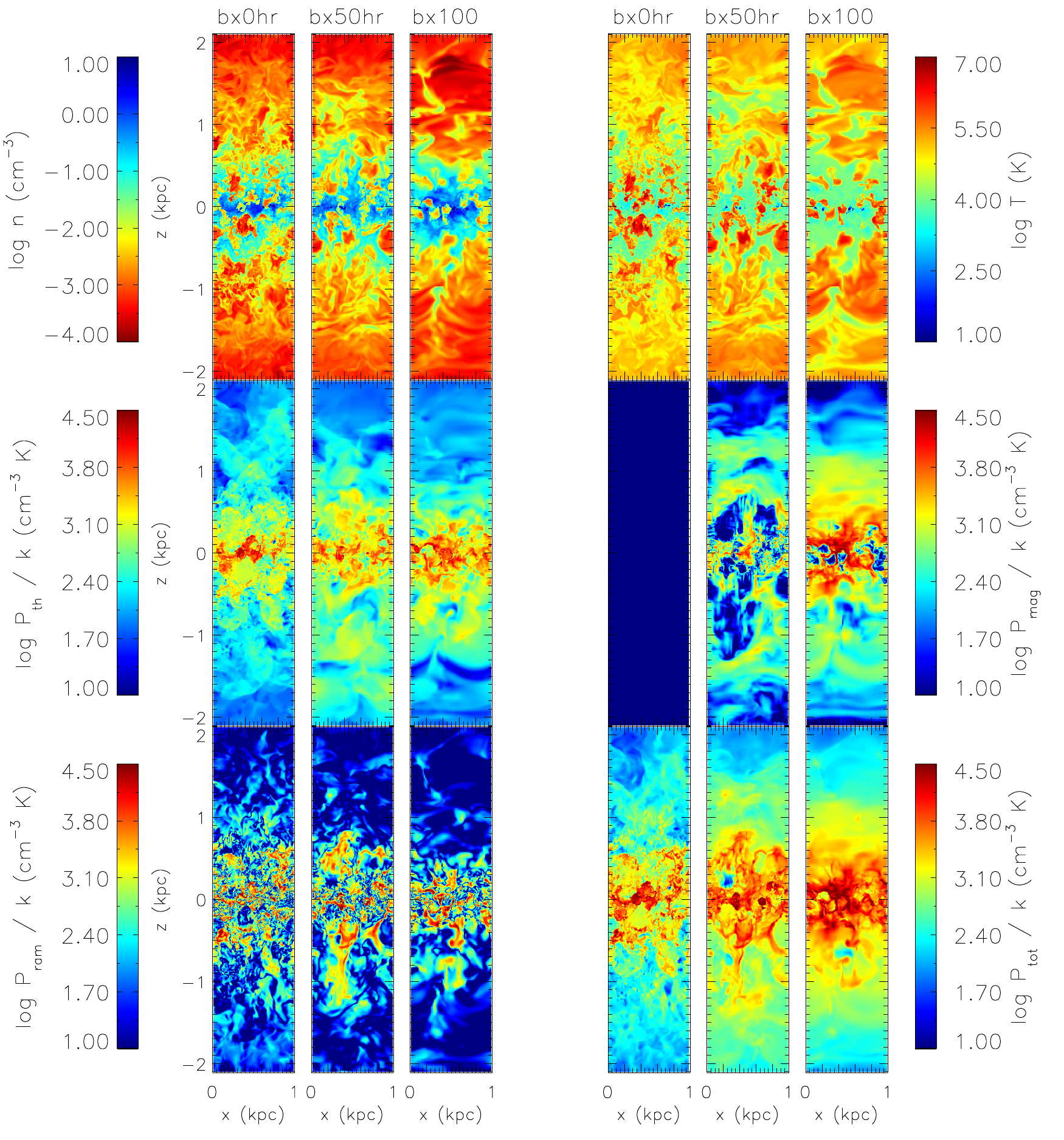}
\caption{Vertical snapshots of density, temperature, thermal pressure, magnetic pressure, vertical ram pressure, and total (thermal plus magnetic plus vertical ram) pressure at $t=340 \Myr$ ($240 \Myr$ for \bxo) from models with a range of initial magnetic field strengths (see Table~\ref{tbl:runs}). Slices from each model along the field are shown.}
\label{fig:images}
\end{\figurestar}

\begin{\figurestar}
\plotone{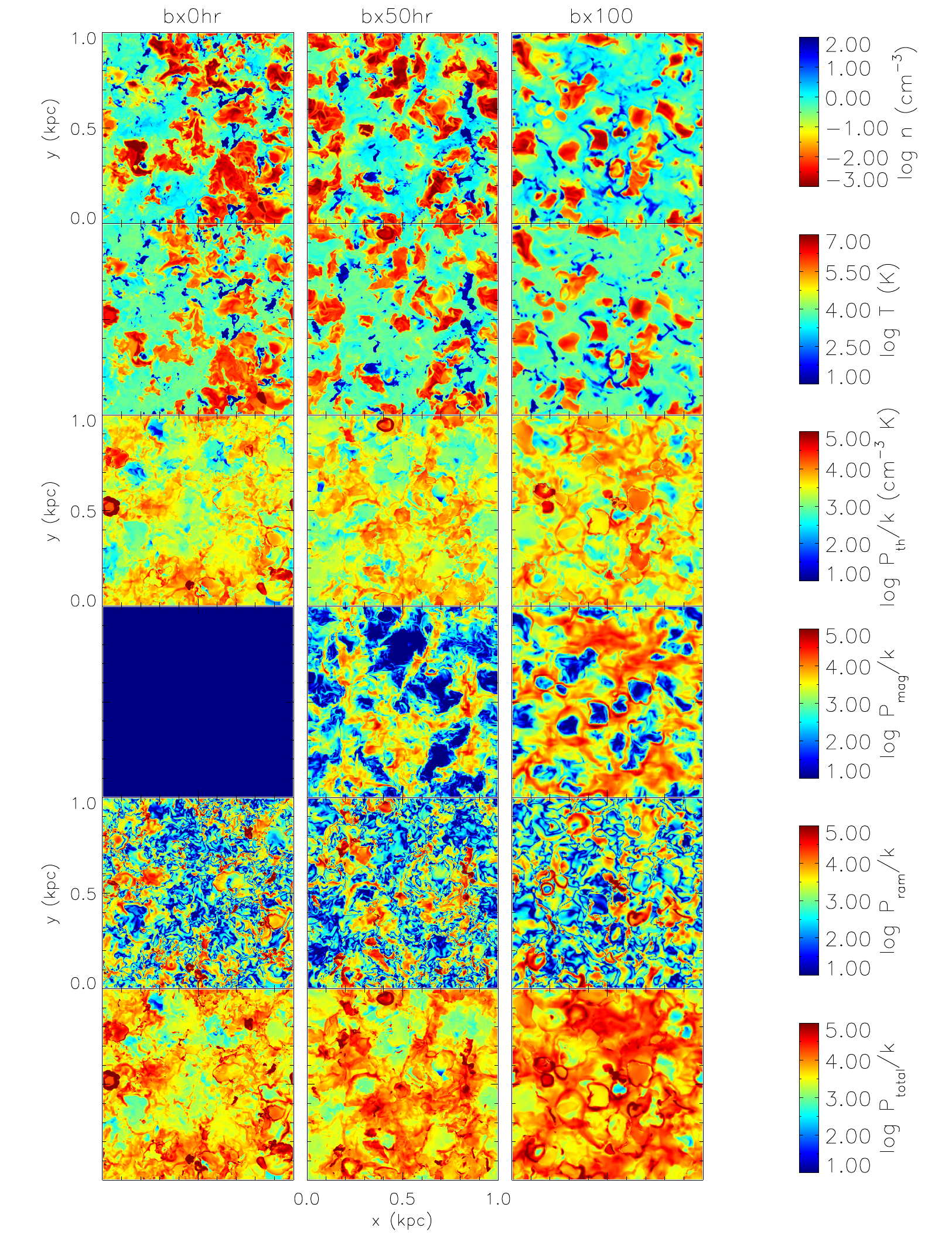}
\caption{Images of the midplane slice ($z=0$) of (top to bottom) density, temperature, thermal pressure, magnetic pressure, vertical ram pressure, and total pressure for a range of initial magnetic field strengths at the same timestamps as in Fig.~\ref{fig:images}. Note that the color scales are different than in Fig.~\ref{fig:images}.}
\label{fig:plane_images}
\end{\figurestar}

\begin{figure}
\plotone{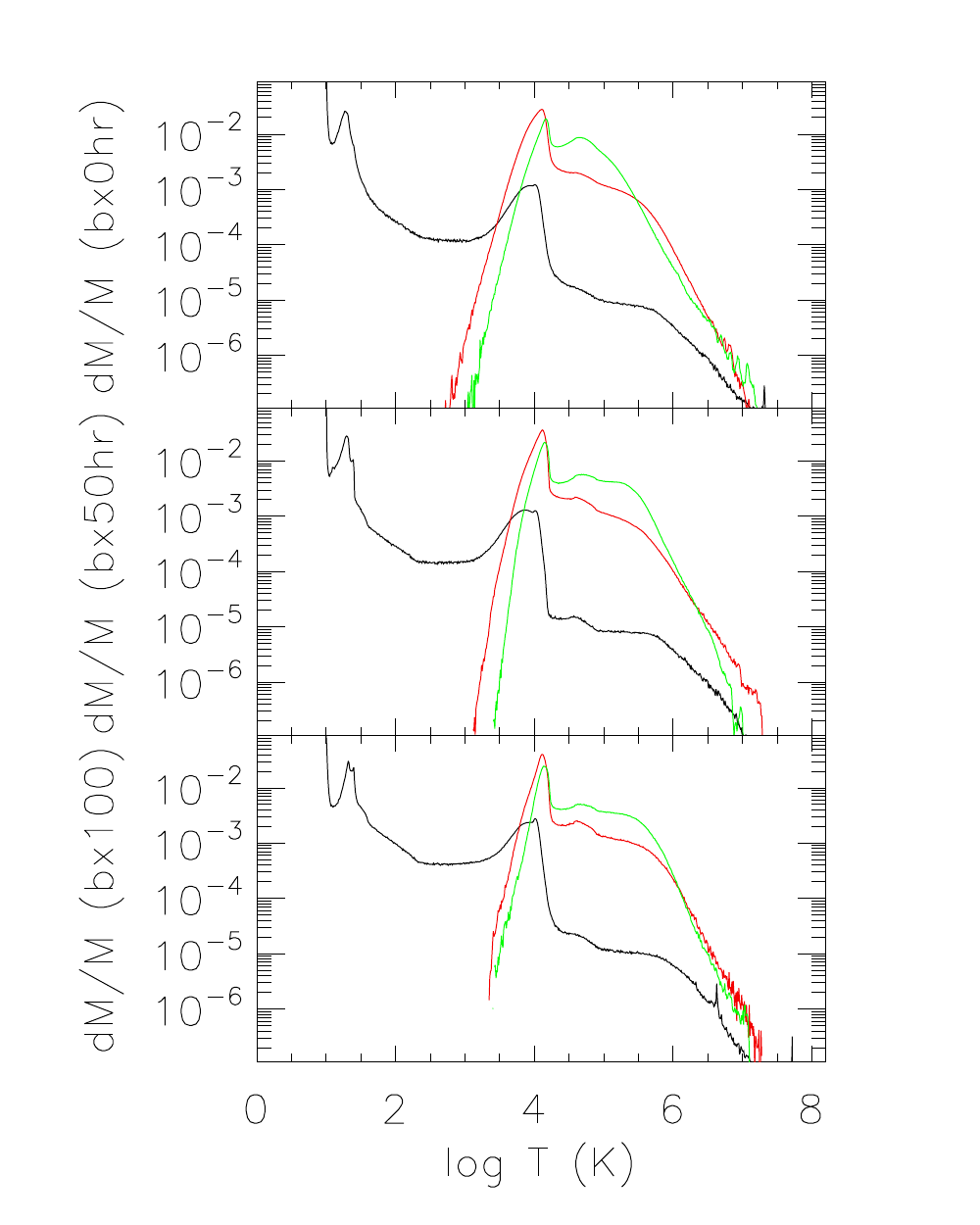}
\caption{Mass-weighted histograms of temperature averaged over $21$ time slices from $t=320-340 \Myr$ ($240-260 \Myr$ for \bxo) in the \bxzhr, \bxfhr, and \bxo\ models (top to bottom). Histograms are calculated within specified height ranges: {\em black lines}: $|z| < 20 \pc$; {\em red lines}: $400 \pc < |z| < 600 \pc$; and {\em green lines}: $800 \pc < |z| < 1200 \pc$. Histograms were calculated with logarithmic bin intervals $d(\log T) = 0.01$.}
\label{fig:hists_dm}
\end{figure}

\begin{\figurestar}
\plotone{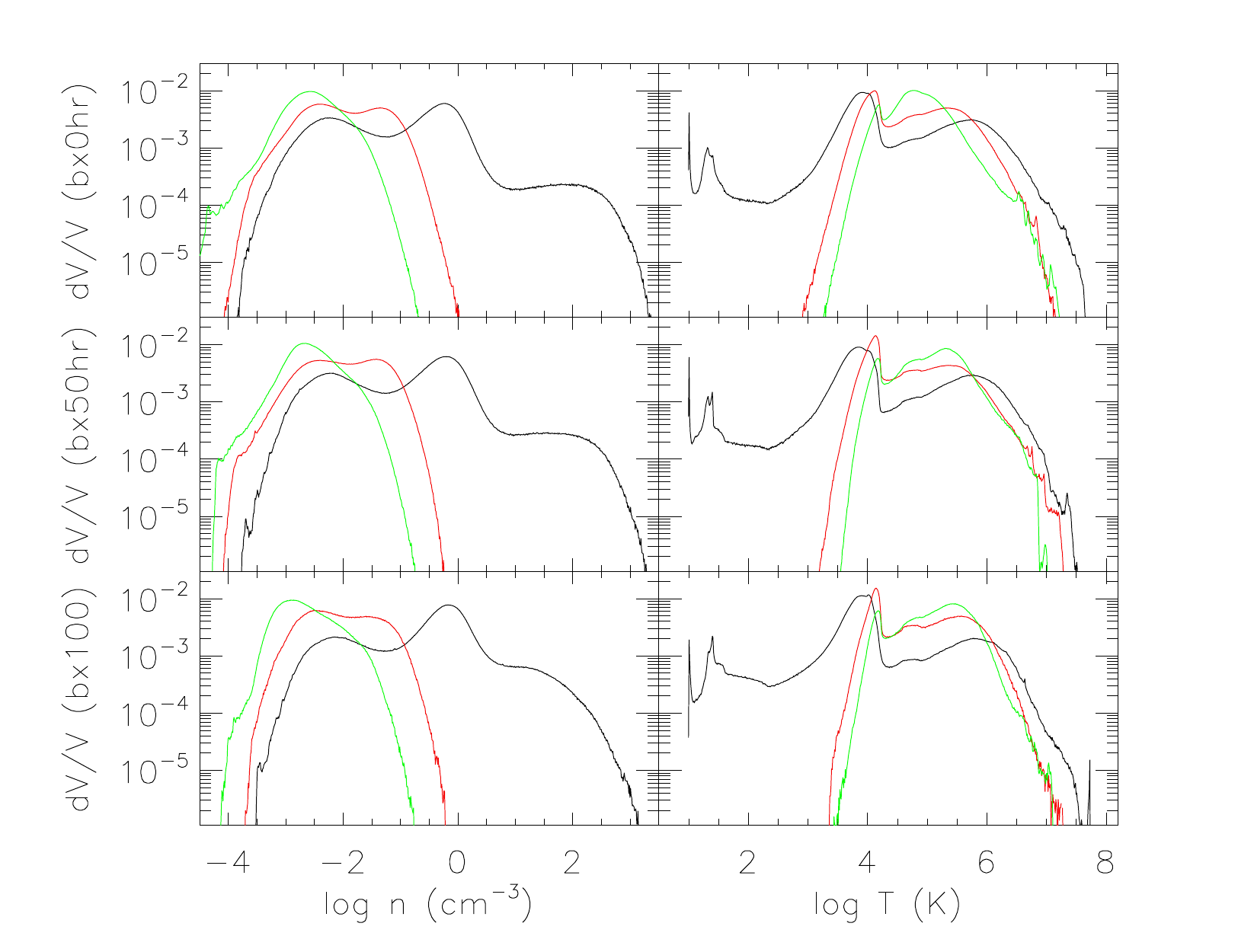}
\caption{Volume-weighted histograms of density and temperature for time slices as in Fig.~\ref{fig:hists_dm}. As in Fig.~\ref{fig:hists_dm}, black lines show the plane, red lines gas at $|z| \sim 500 \pc$, and green lines gas at $|z| \sim 1 \kpc$. Histograms were calculated with logarithmic bin intervals $d(\log n) = d(\log T) = 0.01$.}
\label{fig:hists_dv}
\end{\figurestar}

We show vertical and horizontal slices of a snapshot of the \bxzhr, \bxfhr, and \bxo\ simulations in Figures~\ref{fig:images} and~\ref{fig:plane_images}. The qualitative structure of each of the models is similar, replicating the conclusion of \citet{ab05} that the magnetic field does not prevent the breakout of superbubbles in 3D simulations. Mass-weighted histograms of temperature in the plane (black lines in Fig.~\ref{fig:hists_dm}) show that the classically-defined phases of the ISM at thermally stable temperatures \citep{whm95,wmh03} do exist in our model: there are peaks in the histograms at $\approx 25 \K$ and $\approx 7000 \K$ corresponding to cold and warm thermally stable gas. Comparison of mass- and volume-weighted histograms of temperature (Figs.~\ref{fig:hists_dm} and \ref{fig:hists_dv}) shows that a considerable portion of the volume consists of very low-density hot gas: there is a local maximum in the volume-weighted histogram of midplane temperature at $\approx 10^6 \K$ (black lines in Fig.~\ref{fig:hists_dv}) but no corresponding maximum in the mass-weighted histogram (black lines in Fig.~\ref{fig:hists_dm}).

Figures~\ref{fig:hists_dm} and \ref{fig:hists_dv} also show histograms of temperature and density at $\approx 500 \pc$ (red lines) and $\approx 1 \kpc$ (green lines) above the plane. The distribution of mass is different at these heights. No cold or cool gas is present. The peak corresponding to warm gas is considerably narrower than in the plane, with the peak at $\approx 1.5 \times 10^4 \K$, twice the most probable temperature of warm gas in the plane. There are also local maxima in the transition-temperature portion of the distribution at $6 \times 10^4 \K$, which is a local minimum on our cooling curve (Fig.~1 of JM06) and $2 \times 10^5 \K$, which is thermally unstable and near the maximum cooling efficiency due to \ion{O}{6}. Gas in this transition-temperature regime is normally assumed to lie on interfaces between hot and warm clouds \citep{fws05,sw09} because it is highly thermally unstable. However, transition-temperature gas dominates the volume at $|z| \sim 1 \kpc$ of our model: the transition-temperature gas occupies $> 90 \%$ of the volume between $|z| = 1.5-2 \kpc$ in the \bxfhr\ model, or $> 95 \%$ in the \bxzhr\ model. The transition-temperature gas accounts for a similarly large fraction of the mass at these heights as well. The distribution of temperatures within the transition-temperature regime is affected by the magnetic field, with the peak of the distribution (the green lines in Figs.~\ref{fig:hists_dm} and \ref{fig:hists_dv}) occurring at $6 \times 10^4 \K$ in the unmagnetized run but at $2 \times 10^5 \K$ in the magnetized runs.

To investigate the role of gas in each temperature regime we divide the gas into five temperature regimes informed by the three phase model \citep{mo77}: thermally stable ``cold'' molecular and CNM gas ($T < 200 \K$); thermally unstable ``cool'' gas ($200 \K < T < 5000 \K$), thermally stable ``warm'' gas ($5000 \K < T < 10^{4.2} \K$), thermally unstable ``transition-temperature'' gas ($10^{4.2} \K < T < 10^{5.5} \K$), and ``hot'' gas ($T > 10^{5.5} \K$). We use these definitions for the remainder of this paper.

\begin{figure}
\plotone{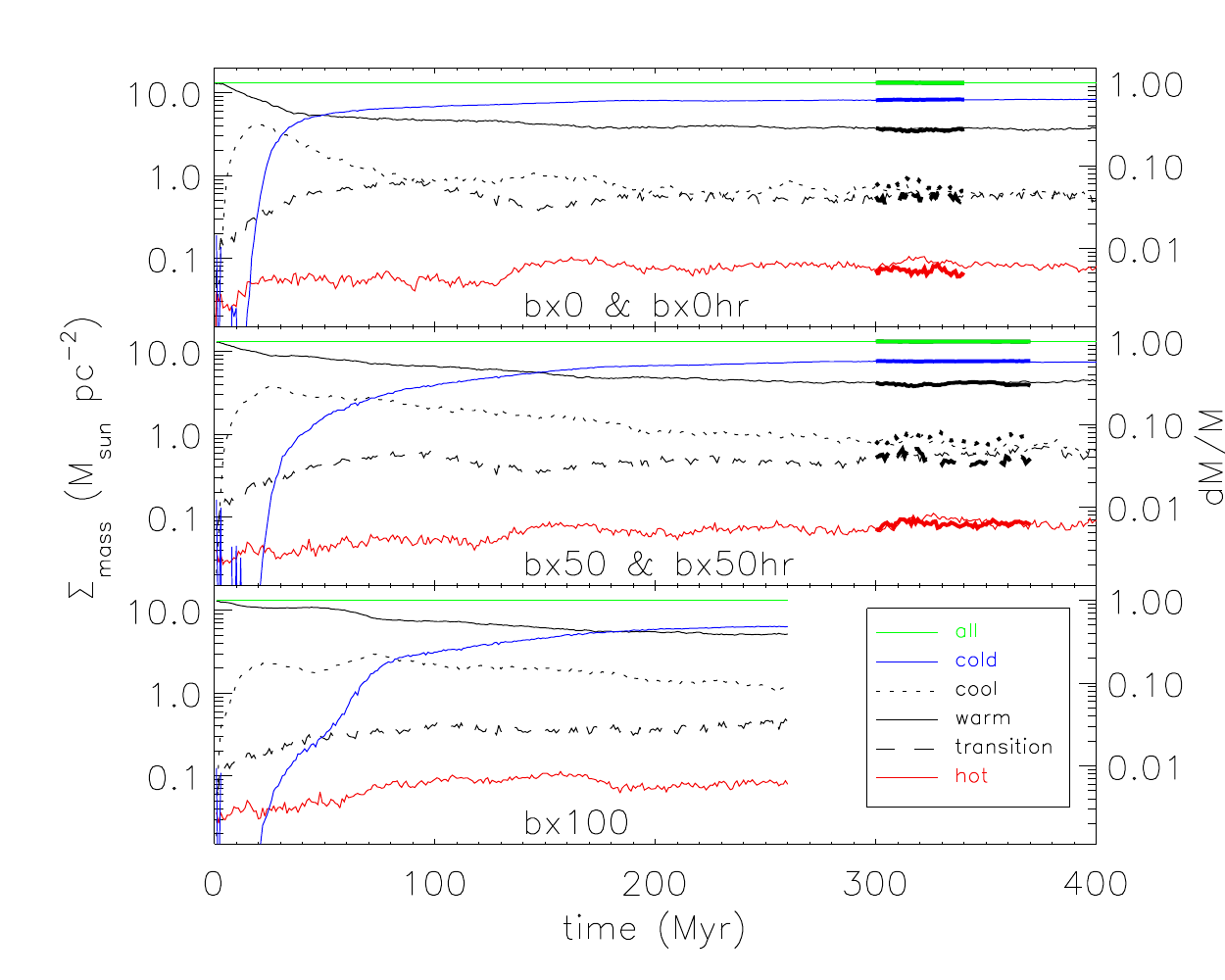}
\caption{Surface mass density as a function of time for the \bxz, \bxf, and \bxo\ models ({\em top to bottom}). The right axis indicates the corresponding mass fraction. The color and line styles are the same in all plots. {\em Blue lines}: $T < 200 \K$. {\em Dotted lines}: $200 \K < T < 5000 \K$. {\em Black solid lines}: $5000 \K < T < 10^{4.2} \K$. {\em Dashed lines}: $10^{4.2} < T < 10^{5.5} \K$. {\em Red lines}: $T > 10^{5.5} \K$. {\em Green lines}: all gas. High resolution versions (\bxzhr\ and \bxfhr) are overplotted (on \bxz\ and \bxf, respectively) with the same line styles and colors but thick lines.}
\label{fig:smd}
\end{figure}

Figure~\ref{fig:smd} shows the evolution of the surface mass density in the entire box for each temperature regime. After $\approx 250 \Myr$, the gas mass in each temperature regime does not change in the \bxz\ and \bxf\ models, indicating that these models have reached a statistical steady-state. The \bxo\ model may have reached a steady-state in the distribution of mass as well but has not run for enough time to verify this conclusion. Figure~\ref{fig:smd} also shows that the formation of cold gas is delayed as the magnetic field strength increases.

Approximately $62 \%$ of the gas mass is in the cold regime (blue lines in Fig.~\ref{fig:smd}) in the \bxz\ model, with $57 \%$ of the mass in the cold phase in the \bxf\ model. The standard deviations of these mass fractions over the time interval $300-400 \Myr$ are $\approx 0.6 \%$. The mass fractions in the high resolution ($1.95 \pc$, shown with thick lines) runs are identical to those in the medium resolution ($3.91 \pc$) ones, suggesting that our results have converged. The steady-state mass fractions in warm gas are $28 \%$ and $31 \%$ for the \bxz\ and \bxf\ models, respectively. After $260 \Myr$, the \bxo\ model has mass fractions of $48 \%$ for the cold and $39 \%$ for the warm gas.

\begin{figure}
\plotone{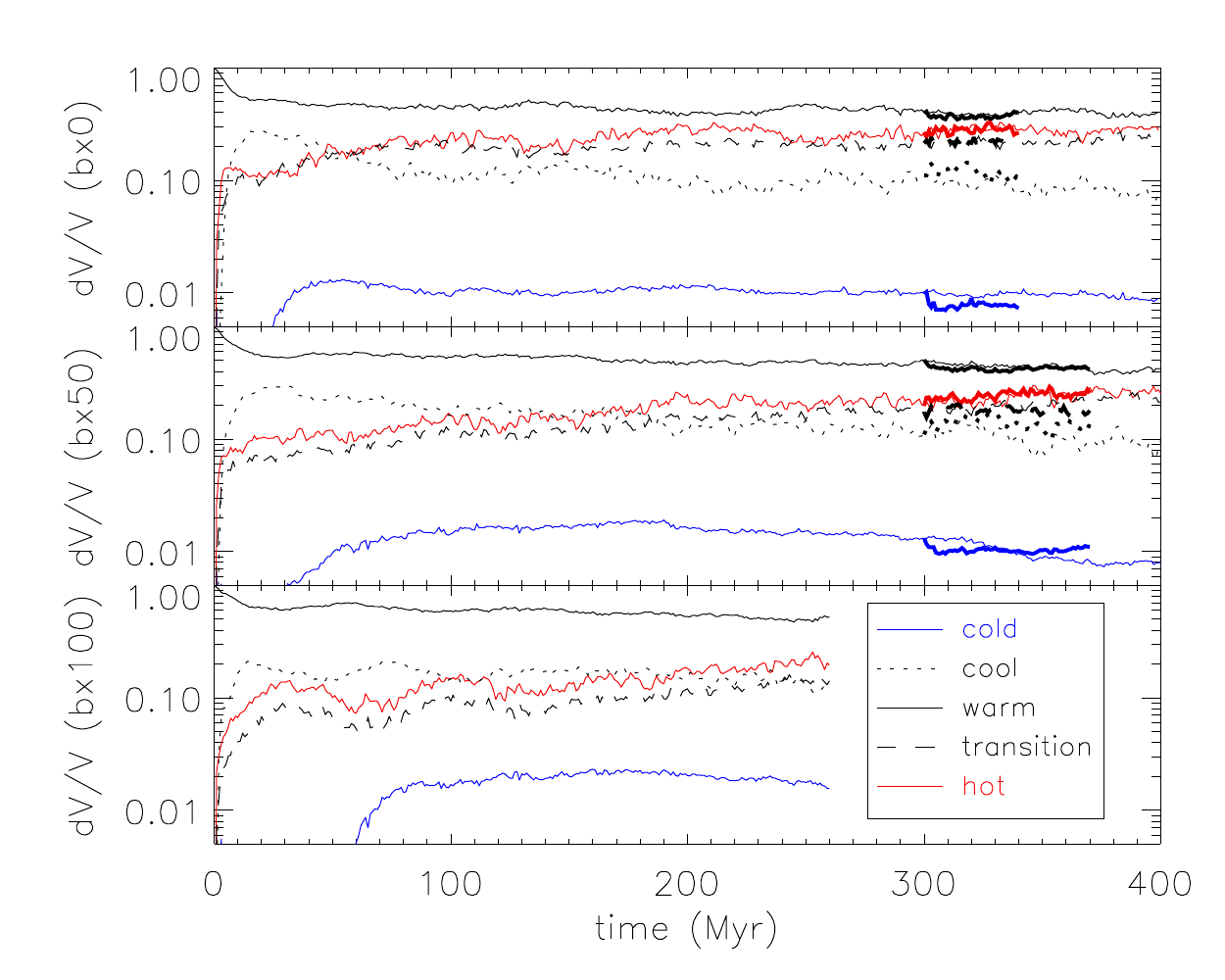}
\caption{Time evolution of volume filling fractions of phases of the ISM, selected by temperature, in the plane ($|z| < 125 \pc$). Plots are shown for models with three different magnetic field strengths (\bxz, \bxf, and \bxo), with the two high resolution models (\bxzhr\ and \bxfhr) shown with thick lines. Line styles are the same as in Fig.~\ref{fig:smd}.}
\label{fig:time_evol}
\end{figure}

We show volume filling fractions of each temperature regime in Figure~\ref{fig:time_evol}. In all cases, the warm, thermally stable gas fills $\approx 36-50 \%$ of the volume near the plane ($|z| < 125 \pc$), more than any other phase; this fraction is highest in the strongly magnetized case and lowest in the unmagnetized case. Thermally unstable cool gas occupies $\approx 15 \%$ of the volume. Hot and transition-temperature gas ($T > 10^{4.2} \K$) fills most of the remaining $30-40 \%$ of the volume, with cool gas occupying $\approx 10\%$ in all cases. This hot gas consists primarily of superbubbles produced by correlated supernovae. Note that the evolved picture seen here as well as in similar, recent work \citep[e.\ g.][JM06]{ab04} is one of a pervasive, warm intercloud medium with embedded supernova remnants and superbubbles. This contrasts with the original three-phase ISM models \citep{cs74,mo77} in which hot gas forms the intercloud medium.

\subsection{Vertical structure} \label{sec:vertical_structure}

\begin{\figurestar}
\plotone{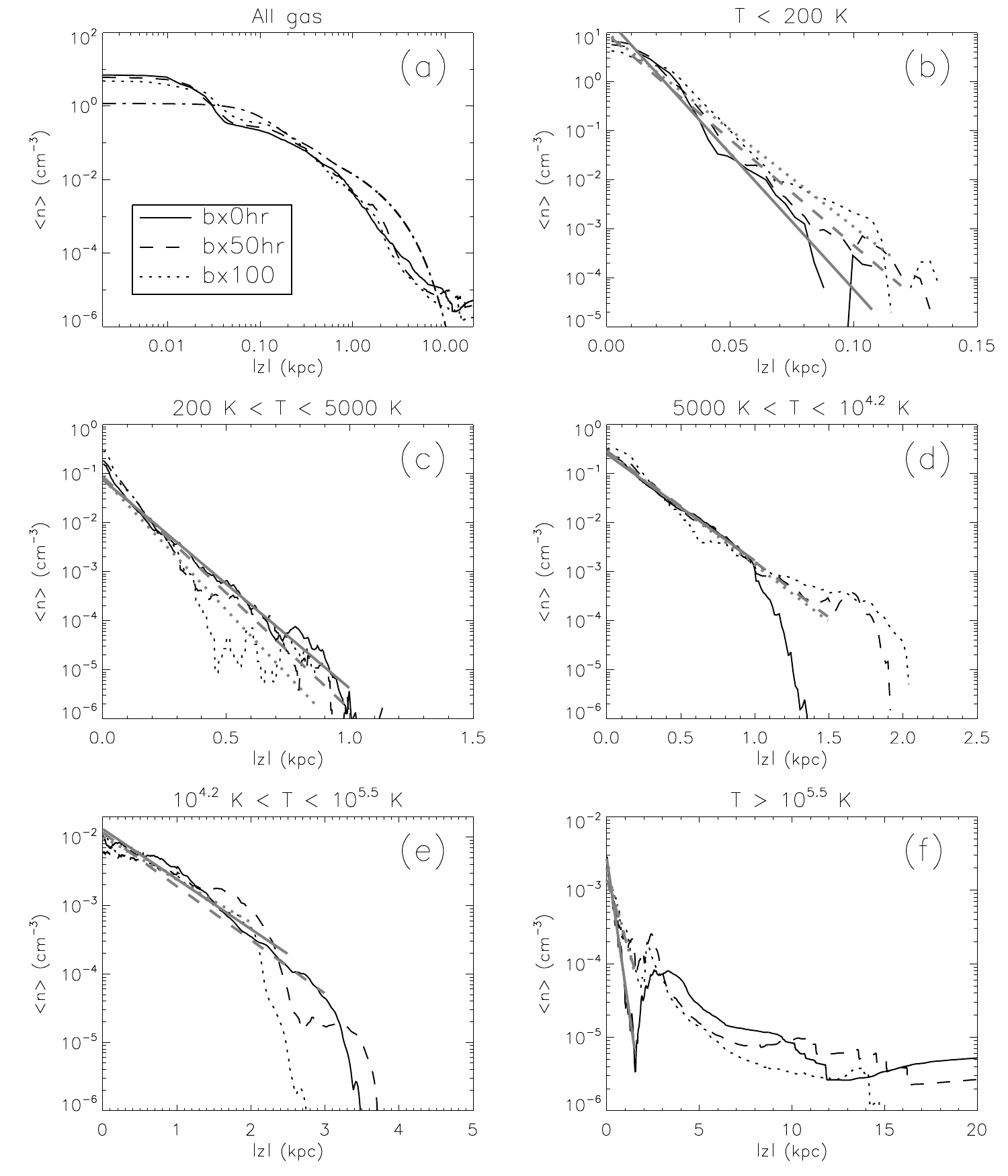}
\caption{Vertical profiles of horizontally-averaged density as a function of height. Profiles are shown for all gas on a log-log scale ($a$) and on a semi-log scale for the cold, cool, warm, transition-temperature, and hot gas ($b-f$). Each plot includes dark lines for the \bxzhr, \bxfhr, and \bxo\ models with data above and below the plane in four, equally-spaced time slices from $330-360 \Myr$ ($230-260 \Myr$ for the \bxo\ model) averaged. Exponential fits are shown with gray lines, with scale heights listed in Table~\ref{tbl:scale_height}. Aspect ratios are different in each plot. The dot-dashed line in panel ($a$) represents the sum of fits of observed $\mathrm{H}_2$, $\mathrm{H}^0$, and $\mathrm{H}^+$ density profiles (see Section~\ref{sec:vertical_structure}).}
\label{fig:profile_by_phase}
\end{\figurestar}

\begin{\figurestar}
\plotone{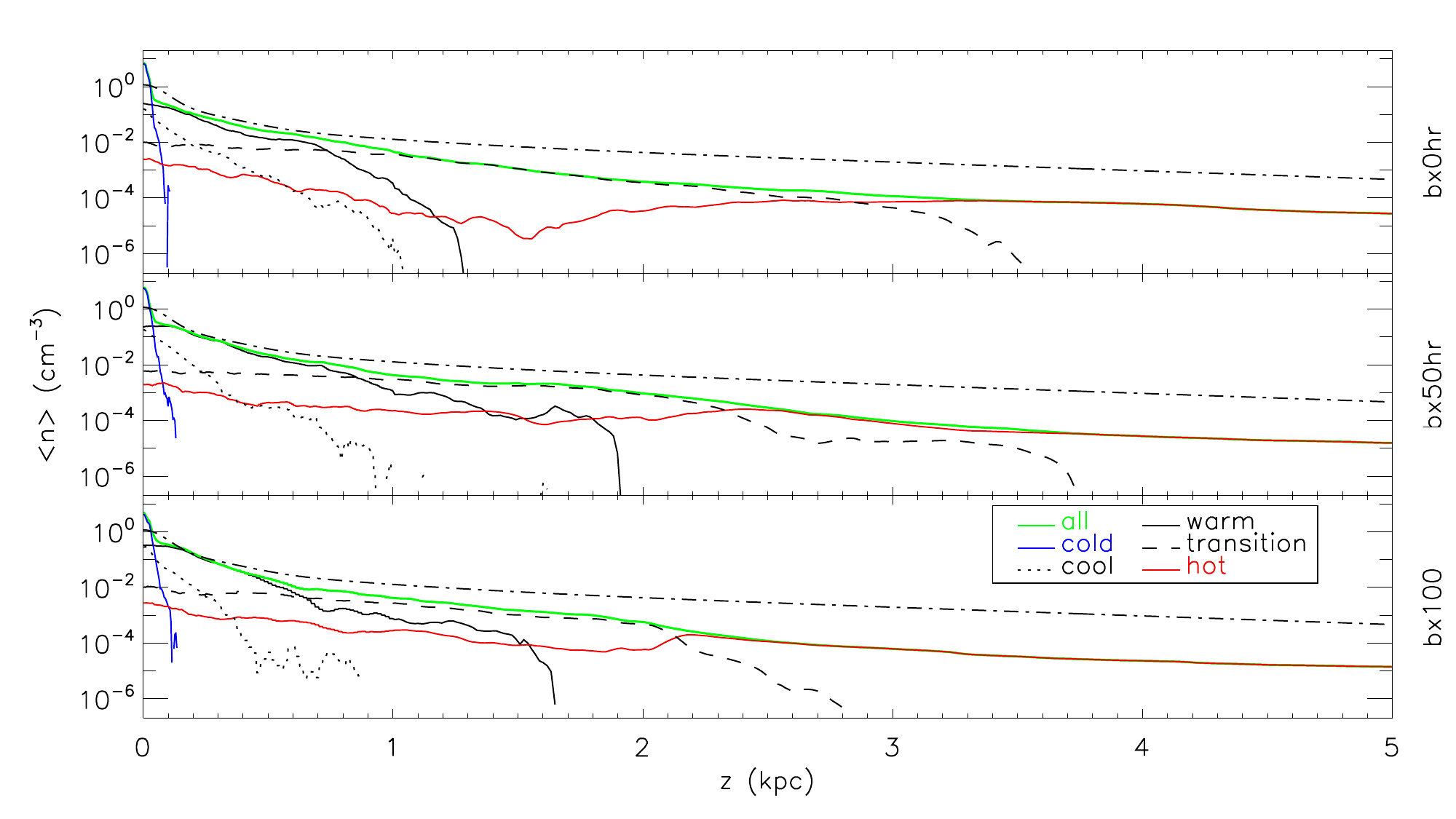}
\caption{Horizontally-averaged density as a function of height, with the \bxzhr, \bxfhr, and \bxo\ models shown separately ({\em top} to {\em bottom}). The data shown and dot-dashed line (representing fits to observed molecular, atomic, and ionized hydrogen profiles) are the same as in Fig.~\ref{fig:profile_by_phase}, while the linestyles depicting the data are the same as in Figs.~\ref{fig:smd} and \ref{fig:time_evol}.}
\label{fig:profile_by_run}
\end{\figurestar}

\begin{\figurestar}
\plotone{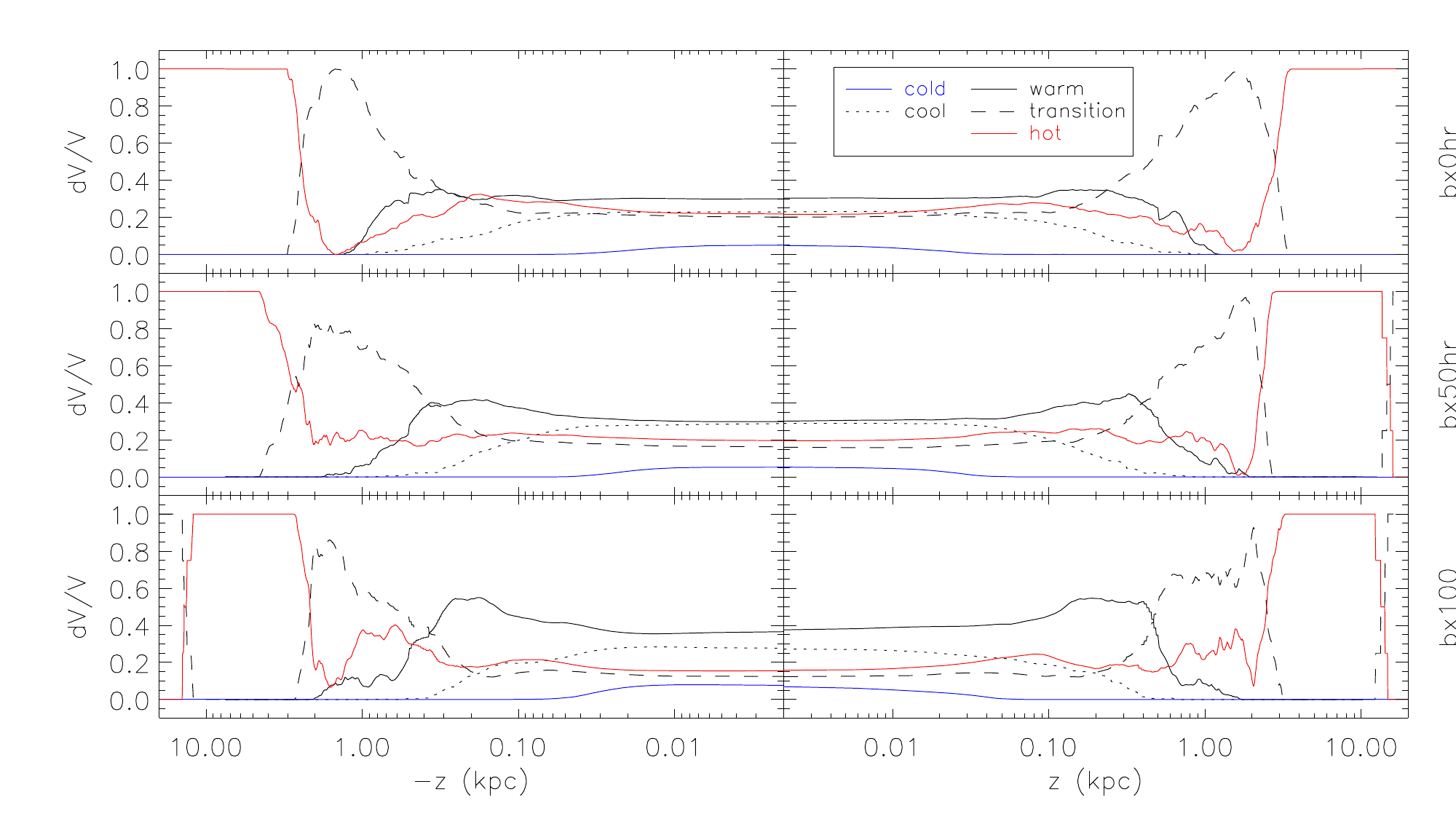}
\caption{Volume filling fraction as a function of height for the \bxzhr, \bxfhr, and \bxo\ models ({\em top} to {\em bottom}) after $340 \Myr$ ($240 \Myr$ for \bxo). Line styles and colors are the same as in Figs.~\ref{fig:smd}, \ref{fig:time_evol}, and \ref{fig:profile_by_run}. Note that the horizontal ($z$) axis scale is logarithmic.}
\label{fig:fillfrac_height}
\end{\figurestar}

We show the space-averaged gas density as a function of height for each temperature regime in Figure~\ref{fig:profile_by_phase}. Vertical profiles of space-averaged density for each of the three models are shown in Figure~\ref{fig:profile_by_run} and of filling fraction in Figure~\ref{fig:fillfrac_height}. For comparison, we also show syntheses of observational density profiles in Fig.~\ref{fig:profile_by_phase}$a$ and \ref{fig:profile_by_run}. The profiles are the sum of a molecular gas profile with a Gaussian scale height of $81 \pc$ \citep{css88,f01}, a \citet{dl90} profile of atomic hydrogen, and an exponential profile with a scale height of $1.4 \kpc$ representing the WIM \citep{sw09}.

In our models, cold gas is confined to a narrow layer about the midplane extending to $|z| \sim 20 \pc$. There is no cold gas at heights $|z| \gtrsim 150 \pc$. Even in the plane ($|z| \lesssim 10 \pc$), the volume filling fraction of the cold gas is very small, $4.2 \%$ and $5.2 \%$ in the \bxzhr\ and \bxfhr\ models, respectively, as seen in the blue lines Figure~\ref{fig:fillfrac_height}. These volume filling fractions are slightly ($\sim 10 \%$) different in the medium resolution models, as can be seen by comparing the thick and thin blue lines in Figure~\ref{fig:time_evol}. The images (top rows of Figure~\ref{fig:plane_images}) show that cold gas primarily exists in filamentary structures in rough thermal pressure equilibrium with the surrounding warm gas.

The scale heights of the warmer phases are larger, with warm gas dominating by volume and mass to $\sim 500 \pc$, above which the transition temperature gas is the dominant phase. The gas is essentially entirely hot above $|z| \approx 3 \kpc$, with the exception of some transition-temperature gas at $|z| > 10 \kpc$ that is above the galactic fountain circulation; this can be seen as the yellow ($T \approx 10^{5.3} \K$) region at the top of Figure~\ref{fig:images_evol} after $210 \Myr$. Although we include a pressure floor due to the intergalactic medium, we do not model the halo self-consistently. Therefore, we place little physical meaning on the transition-temperature gas at $|z| > 10 \kpc$, above the fountain flow.

Two components of hot gas are evident in Figure~\ref{fig:profile_by_phase}$f$. Near the plane, the hot gas vertical density profile is well-described as an exponential distribution that fills $\approx 20 \%$ of the volume. Much of this hot gas has high thermal pressures and low magnetic pressures compared to the surrounding medium (see Section~\ref{sec:pressure} below) and is associated with recent supernovae. The exponential fit to this distribution finds a scale height of $300-500 \pc$, larger than the Type~II supernova scale height of $90 \pc$. At $|z| \sim 1-3 \kpc$, the transition-temperature gas dominates, with a second hot component dominating at heights above the transition-temperature gas.

\subsubsection{Scale heights} \label{sec:scaleheight}

We characterize the vertical extent of phases of the ISM by calculating characteristic heights of gas in each temperature regime using two methods. The numerical values of the scale heights discussed here are best interpreted by comparison to the physical scales we imposed in our numerical setup, which we describe in Section~\ref{sec:simulations} and summarize in Table~\ref{tbl:experimental_setup}, rather than directly to observed values.\notetoeditor{Tables \ref{tbl:experimental_setup} and \ref{tbl:scale_height} should be displayed on the same page if possible.}

First, we fit an exponential distribution,
\begin{equation} \label{eq:exp}
n(z) = n_0 \exp \left( -\frac{|z|}{h_\mathrm{exp}} \right),
\end{equation}
to the density profiles shown in Figure~\ref{fig:profile_by_phase}. Scale heights, $h_\mathrm{exp}$, and space-averaged midplane densities, $n_0$, for each component are listed in Table~\ref{tbl:scale_height}. Each component is fit well by an exponential distribution, though a $\mathrm{sech}^2(z)$ model may fit the data as well without a cusp at $z = 0$ \citep{gbc01} and a Gaussian distribution has traditionally been used for the molecular and cold atomic gas \citep{dl90,f01}. These space-averaged densities are directly proportional to the total mass in the temperature regime at the height $|z|$, although $n(z)$ is not representative of typical densities within the medium because of the low filling fractions and non-Gaussian distribution of densities \citep{hbk08}. In particular, the densities of cold gas are much higher than the space-averaged value shown in Figures~\ref{fig:profile_by_phase} and \ref{fig:profile_by_run} due to the very small volume filling fraction of the cold gas (Figure~\ref{fig:fillfrac_height}). The exponential distributions of the warm and transition-temperature gas are each truncated at densities $\sim 10^{-4} \cucm$ and heights $|z| \sim 1.2-3 \kpc$; the fits in Figure~\ref{fig:profile_by_phase} and Table~\ref{tbl:scale_height} were conducted only at heights below the truncation. The halo has a minimum total density due to the imposed pressure floor modeling the presence of the intragroup medium. This floor leads to hot gas at densities of $\sim 10^{-5.5} \cucm$ at $|z| \gtrsim 5 \kpc$. Due to the multiple components of hot gas, we fit its exponential distribution only to hot gas at $|z| \lesssim 2 \kpc$.

\begin{figure}[tb]
\plotone{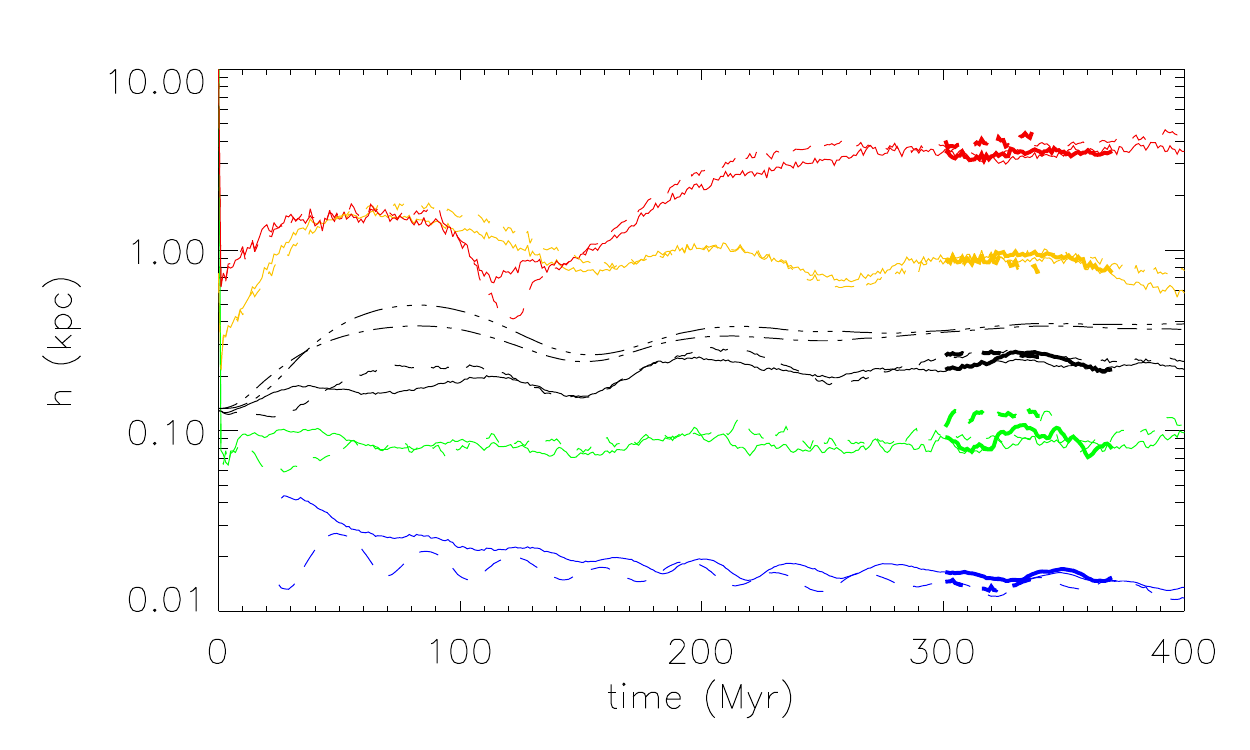}
\caption{Scale heights (equation~\ref{eq:scaleheight}) of components of the simulated ISM as a function of time. Hot gas is shown with red lines, transition-temperature yellow, warm gas black, cool gas green, and cold gas blue; we listed these models in order of decreasing scale height after $200 \Myr$. The \bxf\ and \bxfhr\ models are shown with solid lines, and the \bxz\ and \bxzhr\ models with dashed lines (with the high resolution portion of the runs with thick lines in both cases). The cold gas scale height is not plotted before $25 \Myr$ because of noise before significant amounts of cold gas form. The scale height of all gas is shown with a dot-dashed black line for the \bxf\ model and a triple-dot-dashed black line for \bxz.}
\label{fig:scale_height}
\end{figure}

\begin{deluxetable}{l l r@{.}l}
\tabletypesize{\footnotesize}
\tablecolumns{4}
\tablewidth{0pt}
\tablecaption{Experimental setup}
\tablehead{
	\colhead{Parameter} & \colhead{equation} & \mc{Scale height (pc)} }
\startdata
Gravity ($z \ll z_0 = 180 \pc$)	& (\ref{eq:pot})		& 8&77  \\
Core-collapse Supernovae			&					& 90& \\
Type Ia Supernovae				&					& 325& \\
Diffuse heating					& (\ref{eq:heating})	& 300& 
\enddata
\tablecomments{Characteristic heights of physical processes imposed upon the simulation (described in Section~\ref{sec:simulations}). All of these values are the same for all runs and do not change over the course of the simulations.}
\label{tbl:experimental_setup}
\end{deluxetable}

Second, we calculate the mass-weighted root-mean-square (rms) height of gas in each temperature regime \citep{ko09a}:
\begin{equation} \label{eq:scaleheight}
h = \left( \displaystyle\sum_i \left( \rho_i z_i^2 \right) \middle/ \displaystyle\sum_i \rho_i \right)^{1/2}.
\end{equation}
We performed these estimates with the summations over all zones in each of our temperature regimes; the results are listed in Table~\ref{tbl:scale_height} and plotted as a function of time in Figure~\ref{fig:scale_height}. The scale heights fit to an exponential distribution and those derived from equation~(\ref{eq:scaleheight}) are similar for all components except the hot gas. Though it cannot be directly compared to observational determinations of the scale height of the gas, this second technique provides a less biased characterization of the thickness of the medium than the exponential fits because no truncation of the fit is necessary at low densities.

\begin{\deluxetablestar}{l rrrrr rrrrr}
\tabletypesize{\footnotesize}
\tablecolumns{11}
\tablewidth{0pt}
\tablecaption{Scale heights of simulated ISM components}
\tablehead{\colhead{Model} & \mc{Cold} & \mc{Cool} & \mc{Warm} & \mc{Transition} & \mc{Hot$^\mathrm{a}$} \\
\colhead{} & \colhead{$h$} & \colhead{$n_0$} & \colhead{$h$} & \colhead{$n_0$} & \colhead{$h$} & \colhead{$n_0$} & \colhead{$h$} & \colhead{$n_0$} & \colhead{$h$} & \colhead{$n_0$} }
\startdata
\multicolumn{11}{c}{Exponential fits} \\
\hline
\bxzhr &    8 &  20 &  100 &  0.08 &  200 &   0.3 &  600 &  0.013 &  300 &  0.003 \\
\bxfhr &   10 &   9 &   90 &  0.09 &  190 &   0.3 &  600 &  0.012 &  500 &  0.001 \\
  \bxo &   11 &  10 &   80 &  0.08 &  190 &   0.3 &  700 &  0.010 &  500 &  0.002 \\
\hline
\multicolumn{11}{c}{Mass-weighted rms altitudes (equation~\ref{eq:scaleheight})} \\
\hline
\bxzhr & 14 &&  120 &&  260 &&  800 &&  4000 \\
\bxfhr & 15 &&  100 &&  260 &&  900 &&  3000 \\
\bxo	& 17 &&  100 &&  250 &&  800 &&  2000
\enddata
\tablecomments{Scale heights (in pc) of gas in defined temperature regimes. Fits to an exponential distribution (equation~\ref{eq:exp}) are shown in Figure~\ref{fig:profile_by_phase}. Space-averaged midplane densities $n_0$ (in \cucm) are listed for exponential fits. Mass-weighted rms altitudes are the average from $320-340 \Myr$. \\
$^\mathrm{a}$The exponential distribution of hot gas is fit only to the component near the plane; see text.}
\label{tbl:scale_height}
\end{\deluxetablestar}

\subsection{Magnetic energy} \label{sec:b}

\begin{figure}[tb]
\plotone{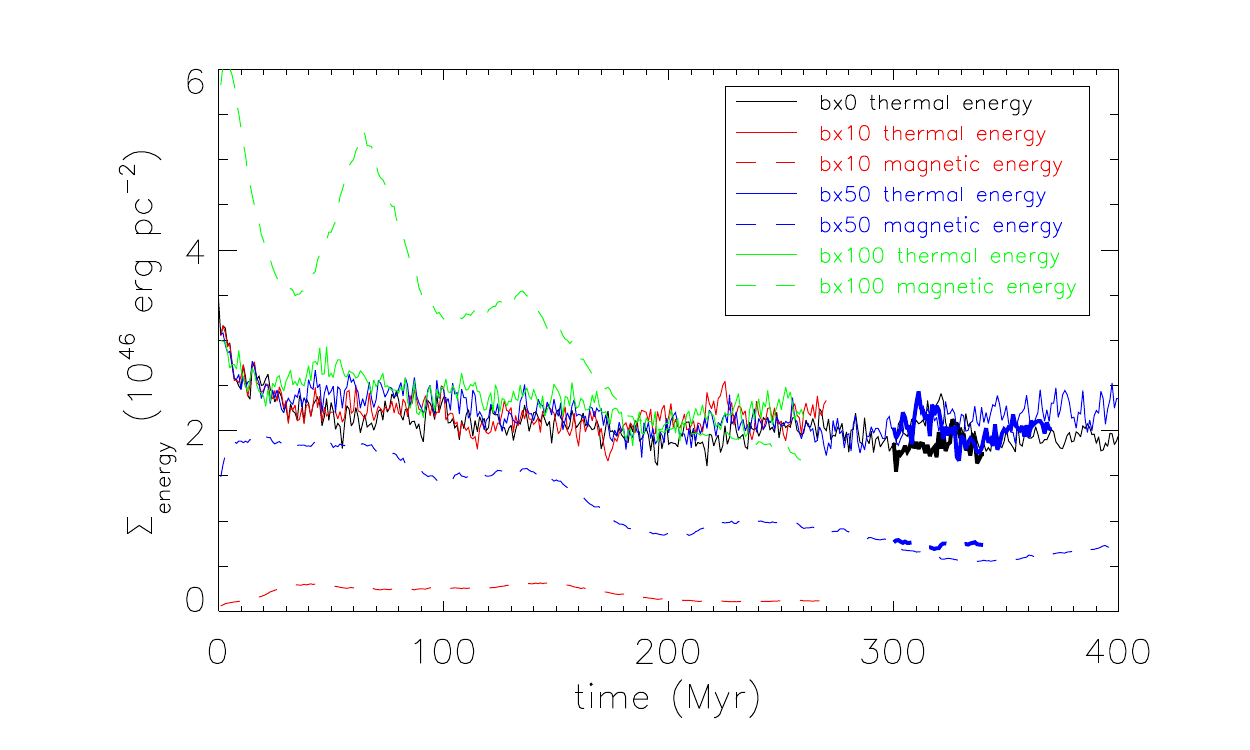}
\caption{Surface magnetic (dashed lines) and thermal (solid lines) energy density as a function of time. The unmagnetized \bxz\ and magnetized \bxt, \bxf, and \bxo\ models are shown with thin lines, while the high resolution counterparts (\bxzhr\ and \bxfhr) are shown with thick lines.}
\label{fig:benergy}
\end{figure}

The evolution of the energy contained in the magnetic field is shown in Figure~\ref{fig:benergy}. In the \bxf\ case, the initial magnetic energy density is approximately half the thermal energy density, whereas the initial magnetic energy density in the \bxo\ case is approximately twice the thermal energy density. In both cases, the thermal energy decreases rapidly and the magnetic energy density increases rapidly in the first $10 \Myr$. This early behavior is during the initial development of a turbulent cascade throughout the plane \citep{bkm04}. After turbulence is established, the magnetic energy in both the \bxf\ and \bxo\ models decreases over time, although there are damped oscillations about the decaying mean with a period of $\sim 70 \Myr$. The field in the \bxf\ model reaches an approximate steady-state at $\approx 1/3$ of the thermal energy density. We have run the \bxo\ model for only $260 \Myr$, and its magnetic field has not reached a steady-state. After $260 \Myr$, the magnetic energy in the \bxo\ has reduced to approximately the same value as the thermal energy in that model.

\subsection{Pressure} \label{sec:pressure}

\begin{figure*}
\plotone{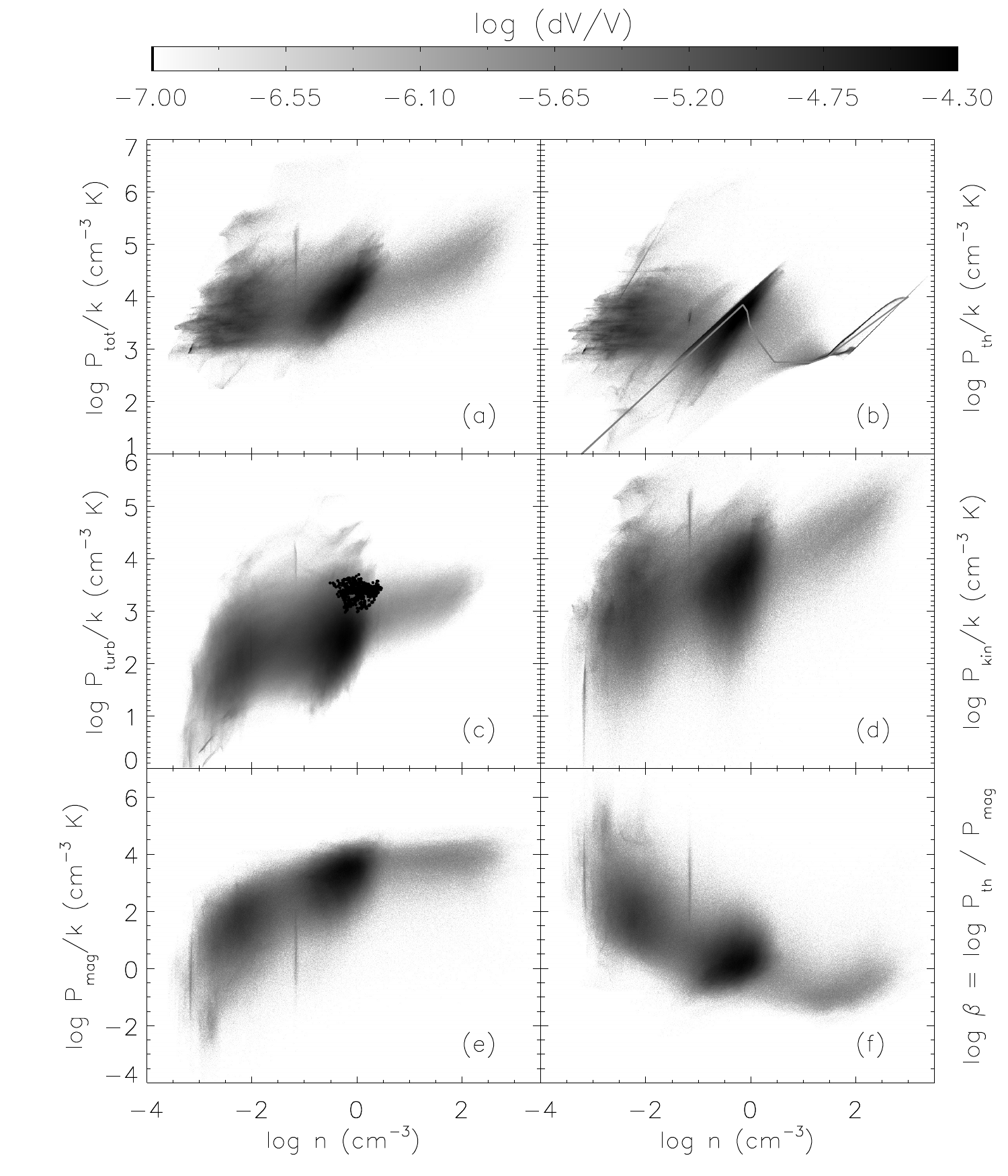}
\caption{Panels $a-e$ represent pressure components ($P/k$) plotted versus density. Panel $a$: Total pressure (sum of thermal, kinetic, and magnetic). Panel $b$: thermal pressure ($nT$). The gray line shows the curve of thermal equilibrium ($n^2 \Lambda = n \Gamma$) for our cooling curve and diffuse heating function at $z=0$ for $T < 2 \times 10^4 \K$. Panel $c$: turbulent pressure ($\langle \rho \rangle \sigma_\mathrm{turb}^2/k$, calculated in boxes $8 \pc$ on a side for $|z| < 125 \pc$). The black dots show turbulent pressure calculated in boxes $125 \pc$ on a side. Panel $d$: kinetic pressure ($\rho |\vec{v}|^2/k$). Panel $e$: magnetic pressure ($|\vec{B}|^2 / 8 \pi k$). Panel $f$ shows the ratio of thermal pressure to magnetic pressure (the plasma $\beta$). All panels except $c$ were calculated for every cell in the range $|z| < 20 \pc$ of a snapshot of the \bxfhr\ model at $340 \Myr$.}
\label{fig:phase}
\end{figure*}

We present images of pressure in Figures~\ref{fig:images} and \ref{fig:plane_images}. We consider three pressures: the thermal ($P_{\mathrm{th}} = nkT$), magnetic ($P_\mathrm{mag} = |\vec{B}|^2 / 8 \pi$), and kinetic pressure ($P_\mathrm{kin} = \rho |\vec{v}|^2$). Plots of these three pressures in the plane as well as their sum are shown as a function of density (phase diagrams) in Figure~\ref{fig:phase}. In the cold (highest density) gas, the non-thermal pressures dominate, with the kinetic pressure being the largest component. The rms velocity in the $T < 200 \K$ gas is $2.5 \kms$, with typical sonic Mach numbers of $\sim 5$. The three pressure components are closest to equipartition in the warm gas ($n \approx 1 \cucm$), although the median magnetic pressure in the warm gas is $\approx 40 \%$ of the median thermal or magnetic pressure. The magnetic pressure is very small in the hot gas, where thermal and kinetic pressures are comparable to each other.
The total pressure in the cold gas is typically a factor of $\sim 3-4$ larger than that found in the warm and hot gas. A small amount of hot gas has thermal pressures $\pth/k > 10^5 \cucm \K$, much larger than the total pressures in cold clouds; this overpressured hot gas is typically associated with supernovae which occurred within the last Myr.

In Figure~\ref{fig:phase}$c$, we show an estimate of the turbulent pressure, calculated following equations~($2-4$) of JMB09. In a series of boxes $8^3 \pc^3$ in size, we calculate the mass-weighted rms velocity dispersion, $\sigma_\mathrm{turb}$, in the frame of the center of mass of the box. The turbulent pressure is $\langle \rho \rangle \sigma_\mathrm{turb}^2$, where $\langle \rho \rangle$ is the mean mass density within the box. We fit a power law function to the relationship between density and turbulent velocity dispersion within $8 \pc$ boxes, $\sigma_\mathrm{turb} \propto \rho^{-\alpha}$. We find that a power law index $\alpha = -0.35$ best fits the \bxfhr\ model, shallower than the value $\alpha = -1/2$ reported by JMB09; we find a similar ($\alpha = 0.35-0.40$) scaling with larger boxes, $31 \pc$ on a side, and in the unmagnetized model. Equivalently, instead of the isobaric behavior reported by JMB09, we find that the turbulent pressure increases with density: median turbulent pressures are $\log(\Pturb{8} / k) = 2.1$ (in $\cucm \K$ units) for the hot gas (defined here by density as $n < 10^{-1.5} \cucm$), $2.5$ for the warm gas ($0.1 \cucm < \langle n \rangle < 3 \cucm$), and $3.2$ for the cold gas ($\langle n \rangle > 10 \cucm$). This trend is in line with that found in the kinetic pressure (Fig.~\ref{fig:phase}$d$). For comparison, the turbulent pressure calculated in $125^3 \pc^3$ boxes is $\log(\Pturb{125} / k) = 3.4$. All $125^3 \pc^3$ boxes have mean densities $\sim 1 \cucm$, so we cannot calculate the turbulent pressure in cold or hot gas on large size scales. Because these estimates capture only the turbulent pressure on size scales of $\sim 8 \pc$, we use the kinetic pressure to calculate the total pressure and for direct comparisons to the thermal and magnetic pressures across widely-varying densities.

\begin{figure}
\plotone{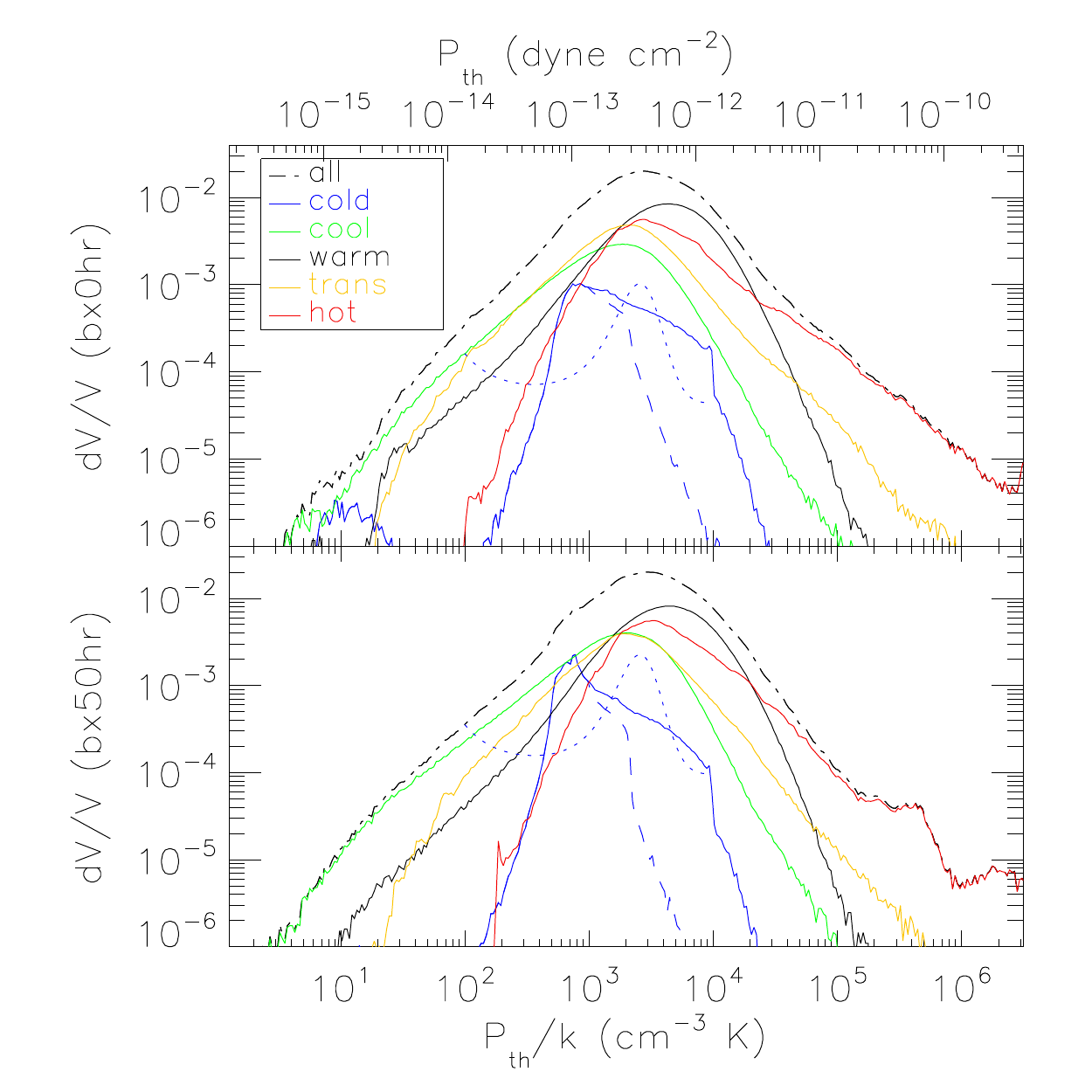}
\caption{Volume-weighted histograms of thermal pressure. The \bxzhr\ ({\em top panel}) and \bxfhr\ ({\em bottom panel}) models in the range $|z| < 20 \pc$ are shown averaged over 21 time slices from $320-340 \Myr$. Gas in each temperature range is identified by color in the legend. The dashed blue line shows only cold gas from our simulation with $n < 100 \cucm$. The distribution identified in \ion{C}{1} data for the CNM by \citet{jt11} is shown with the blue dotted line, with the maximum normalized to the maximum value of $dV/V$ for the cold gas distribution in our simulation. (The dotted line is not otherwise fit to our results.) Pressures are labeled in both $P/k$ units (bottom axis) and cgs units (top axis). Histograms were calculated with logarithmic bin intervals $d(\log P/k) = 0.01$.}
\label{fig:pressure_temp}
\end{figure}

Considering the thermal pressure by temperature reveals the relative importance of the pressure components in each phase of the ISM. In Figure~\ref{fig:pressure_temp}, we show histograms of thermal pressure in the plane for the cold, cool, warm, transition-temperature, and hot gas. The highest thermal pressures are found in hot gas, while the lowest are found in cool gas. The range of pressures found in cold gas is smaller, with $2.5 \lesssim \log (\pth/k) \lesssim 4$. The distribution of each $T>200 \K$ temperature regime is roughly symmetric about the most probable value of $\log \pth$, while the distribution of pressure for the cold gas is highly asymmetric, with a peak at $\pth / k \approx 10^3 \cucm \K$ but pressures extending to $10^4 \cucm \K$. Cool, warm, and transition-temperature gas are all found with thermal pressures both above and below those found in the cold gas. We note that \citet[Fig.~6]{ab05} found the cold gas to have a lower thermal pressure than the other phases.

\subsection{Vertical support} \label{sec:vertical_support}\notetoeditor{Placing Figures 16 and 17 on the same page would be nice.}

\begin{figure*}
\plotone{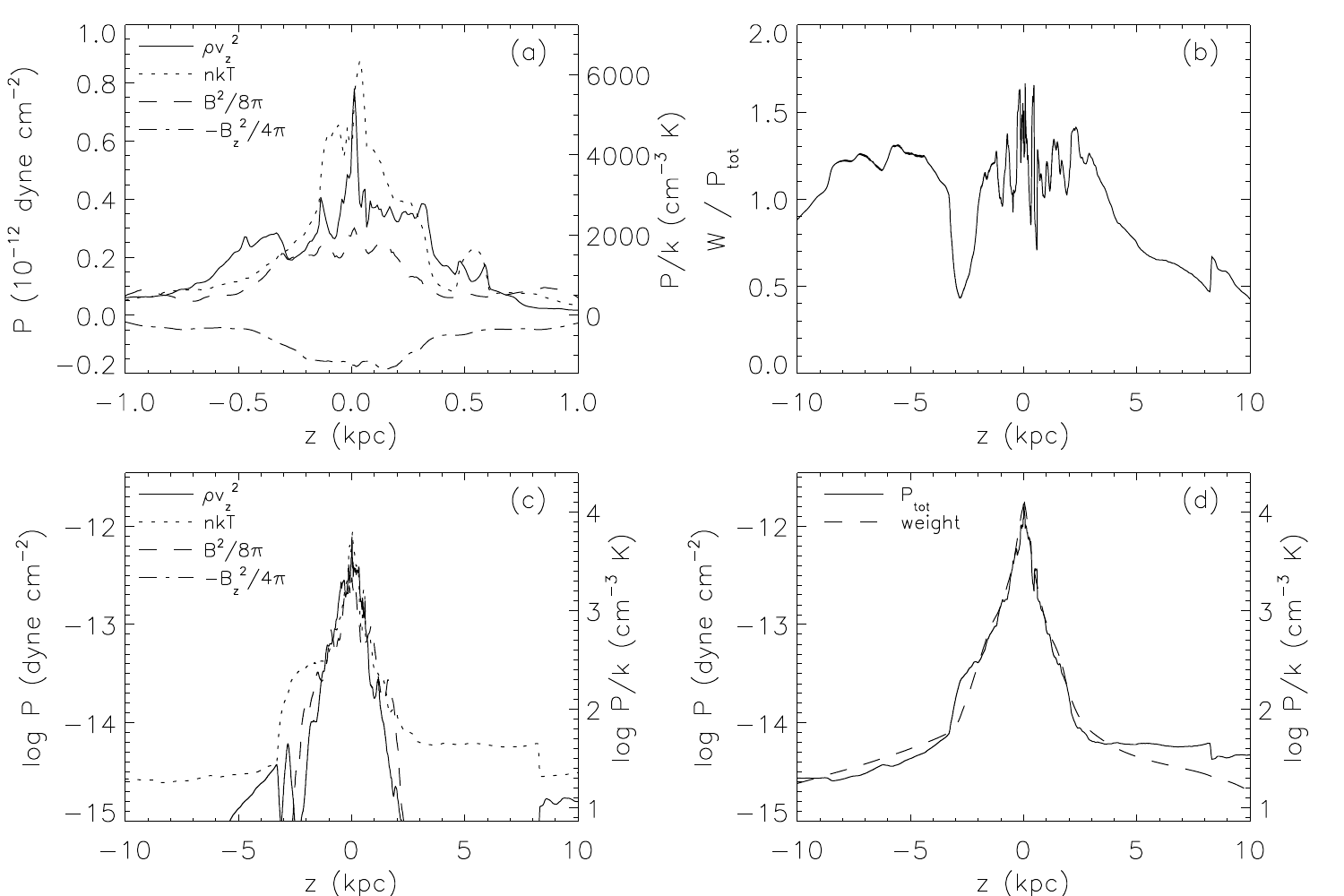}
\caption{{\em Panel $a$:} kinetic ($\rho v_z^2$), thermal ($nkT$), magnetic ($B^2 / 8\pi$) and magnetic tension ($-B_z^2 / 4 \pi$) pressures for the $t=340 \Myr$ snapshot of the \bxfhr\ model at $|z| < 1 \kpc$. Values are space-averaged over $x$ and $y$ for each height $z$. {\em Panel $b$:} Ratio of total weight to total pressure. {\em Panel $c$:} Same as panel $a$ but with a logarithmic vertical axis and expanded to show $|z| < 10 \kpc$; the magnetic tension is omitted because of its negative value. {\em Panel $d$:} Total pressure $P_\mathrm{tot}(z)$ (dashed line) and weight $W(z)$ (solid line, as defined in equation~[\ref{eq:weight}]) as a function of height.}
\label{fig:hydrostatics_bx50}
\end{figure*}

\begin{figure*}
\plotone{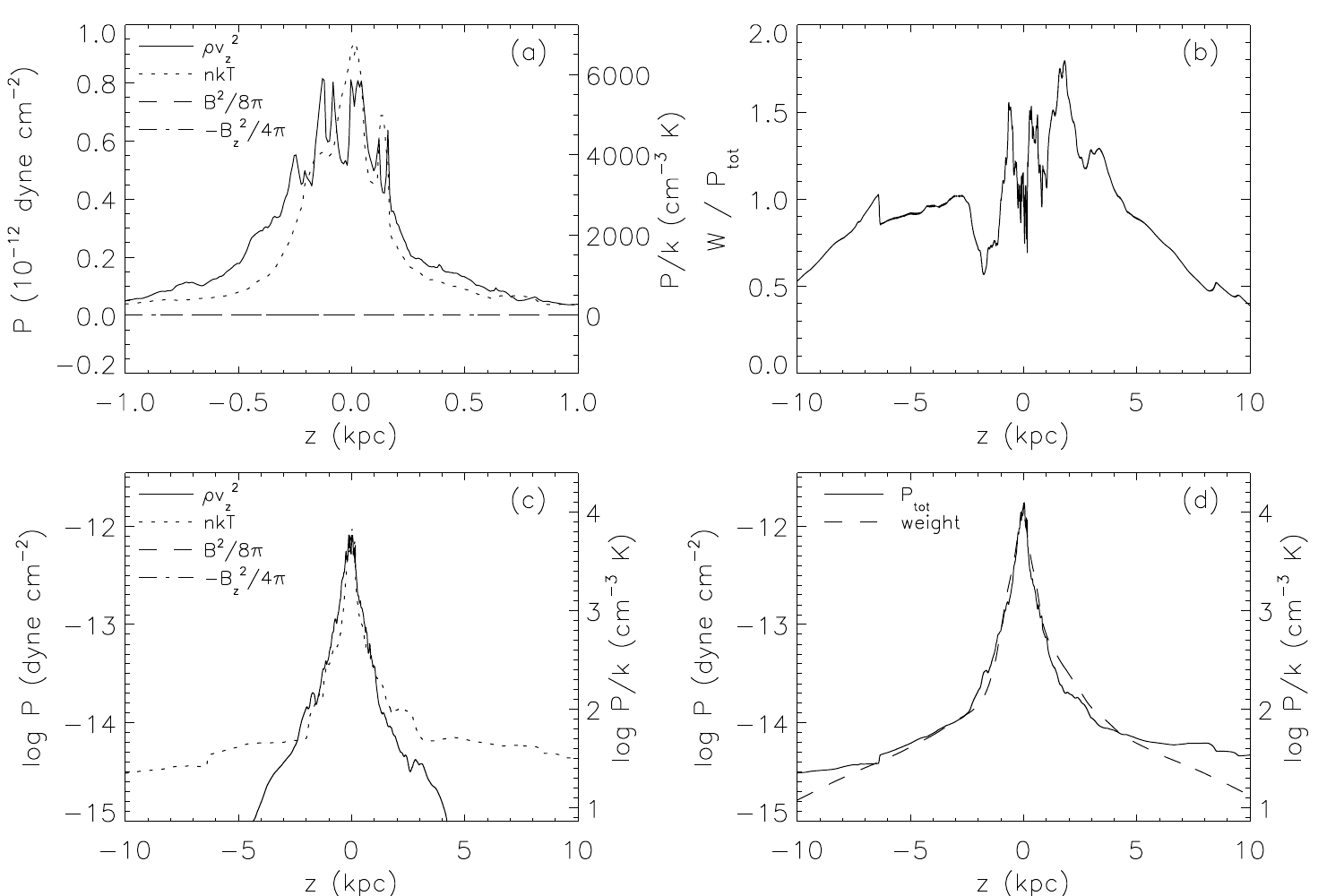}
\caption{Same as Figure~\ref{fig:hydrostatics_bx50} but for the $t=340 \Myr$ snapshot of the unmagnetized \bxzhr\ model.}
\label{fig:hydrostatics_bx0}
\end{figure*}

What provides the vertical support for the stratified ISM in our models? To address this question, we evaluate the roles of vertical kinetic pressure, thermal pressure, magnetic pressure, and magnetic tension \citep{po07}:
\begin{equation}
P_\mathrm{tot}(z) = \left\langle \rho v_z^2 \right\rangle + \left\langle nkT \right\rangle + \frac{\left\langle |\vec{B}|^2 \right\rangle}{8 \pi} - \frac{\left\langle B_z^2 \right\rangle}{4 \pi},
\end{equation}
where the brackets denote volume averaging over $x$ and $y$ positions at fixed $z$. The contributions of these vertical pressure components with height are shown for a snapshot of the \bxfhr\ model in Figure~\ref{fig:hydrostatics_bx50}$a$ and $c$ and for a snapshot of the \bxzhr\ model in Figure~\ref{fig:hydrostatics_bx0}$a$ and $c$.
If the gas is in hydrostatic equilibrium, the pressure at a height $z$ balances the weight $W(z)$ of the gas above $z$. Because all components of the pressure at the top of our box ($|z| = 20 \kpc$) are negligible, this condition would imply \citep{po07}
\begin{equation} \label{eq:weight}
P_\mathrm{tot}(z) = \int_z^{z_\mathrm{max}} g \, \rho \, dz \equiv W(z).
\end{equation}

We compare the weight and pressure of the ISM in panels $b$ and $d$ of Figures~\ref{fig:hydrostatics_bx50} and \ref{fig:hydrostatics_bx0}. As required by the identical surface mass densities and similar scale heights of the magnetized and unmagnetized models (see Section~\ref{sec:scaleheight}), the total pressure profiles are similar in the two models. In both cases, the kinetic and thermal pressure terms are each significant, with the thermal pressure larger than the kinetic pressure at the midplane. In the magnetized model, the magnetic pressure is smaller than either the kinetic or thermal terms at most heights. The total, large-scale impact of the magnetic field is smaller still because the sum of the magnetic tension and magnetic pressure is very small: the maximum of $|\vec{B}(z)|^2 (8 \pi)^{-1} - B_z^2(z) (4 \pi)^{-1}$ is $<10 \%$ of the maximum of $P_\mathrm{tot}$. Equivalently, the mean value of the horizontal field, $B_x^2 + B_y^2$ is approximately equal to the vertical field, $B_z^2$.

%%%%%% DISCUSSION %%%%%%
\section{Discussion} \label{sec:sn_mhd_discussion}

Supernova-driven simulations produce an ISM with mass and height distribution among temperature regimes of the ISM broadly similar to observed values. This suggests that dynamical pressure is the principal driver of vertical stratification of the medium and that supernovae provide the bulk of the dynamical energy.

\subsection{Phases and diffuse heating} \label{sec:phases_heating}

We find strong peaks for warm and cold gas in mass- and volume-weighted histograms of temperature corresponding to temperature regimes that are thermally stable as determined by the cooling curve (Figs.~\ref{fig:hists_dm} and \ref{fig:hists_dv}), similar to the magnetorotational instability-driven simulations of \citet{po07}. This contrasts with the results reported by \citet{ab04,ab05}, in which PDFs of temperature are essentially flat, while \citet{kbs99} find weaker but still significant peaks in their simulations (which extend to only $|z| = 1 \kpc$ and therefore do not track the galactic fountain flow). The difference is likely due to the adopted diffuse heating rate. As noted by JM06, we have chosen a diffuse heating rate too small to maintain the hydrostatic equilibrium initial condition absent supernovae because, in reality, the pressure support for hydrostatic equilibrium is maintained by the combination of photoelectric heating (modeled by our diffuse heating term), heating due to supernovae, and nonthermal pressure. \citet{ab04,ab05}, following the example of \citet{mmn89} and \citet{mf99}, used a diffuse heating term that is sufficient to maintain hydrostatic equilibrium by itself. Their diffuse heating term is therefore $18$ times larger than that used here. Our diffuse heating rate does allow $\approx 12 \%$ of the total gas mass to cool to the minimum allowed temperature, $10 \K$, as can be seen in the peak in histograms of temperature (Figs.~\ref{fig:hists_dm} and \ref{fig:hists_dv}). Of the cells in the simulation that cool to the minimum temperature, $85 \%$ (accounting for $95 \%$ of the mass at the minimum temperature) have a density $n > 100 \cucm$.

Proper treatment of the diffuse heating requires the inclusion of radiative transfer to allow a heating rate that is not constant in both time and space. \citet{phm03} argued that the FUV radiation field intensity in the solar neighborhood varies by factors of $2-20$ on timescales of $\sim 100 \Myr$. Furthermore, the coldest gas will be heated by ionization from local star formation \cite[e.g.][]{m02} in ways that we do not model here. Therefore, a detailed model of the distribution of temperatures into phases must account for these effects. 

\subsection{Filling fraction of thermally unstable atomic gas}

Comparisons of \ion{H}{1} emission and absorption data find that gas in the range $500 \K < T < 5000 \K$ accounts for $\gtrsim 50\%$ of the WNM column density \citep{ht03}. We conduct two crude estimates of the fraction of WNM gas in our simulations that is thermally unstable. If we assume that $\sim 30 \%$ of the warm gas in our simulation is ionized \citep{f01} while the cool gas is all neutral, thermally unstable gas accounts for $\sim 40 \%$ of the mass in the range $200 \K < T < 10^{4.2} \K$ in our magnetized and unmagnetized simulations. This estimate is relatively insensitive to the chosen maximum temperature considered ``thermally stable''; we use the cutoff of $5000 \K$ used by \citet{ht03}. Alternatively, $50 \%$ of the mass in the $500 \K < T < 8000 \K$ range is cooler than $5000 \K$; this value is the same in our magnetized and unmagnetized models. We note that \citet{ab05} found a $60 \%$ unstable WNM fraction using this estimate, although (as noted above) their gas is roughly evenly distributed across all temperatures whereas ours is concentrated at temperatures $T \gtrsim 3000 \K$.

Photoionization modeling promises to allow a cleaner discrimination between WNM and WIM gas. We also plan to construct synthetic \ion{H}{1} observations through our simulations to refine this estimate. Although diffuse ionized gas is rarely observed at temperatures $< 6000 \K$ \citep{hrt99,mrh06}, gas at $T < 6000 \K$ {\em in our simulations} could be ionized: photionization heating, the dominant heating source in the WIM at densities $n \gtrsim 0.1 \cucm$ \citep{rc92,rht99}, would increase the temperature of such gas.

\subsection{Thermal pressure of CNM and molecular gas} \label{sec:pressure_discussion}

\citet{jt11} have compiled a set of thermal pressures in CNM gas using {\em Space Telescope Imaging Spectrograph} \ion{C}{1} observations. They find that the CNM pressure is lognormally distributed when weighted by mass. Their estimated volume-weighted distribution, calculated assuming that the CNM gas is in thermal equilibrium (effective $\gamma = 0.7$), is plotted with the dotted blue line in Figure~\ref{fig:pressure_temp}. The highest thermal pressure in cold gas in the model is higher than observed in the CNM while the peak of the modeled distribution is at lower pressure than most of the observed CNM pressures. Although we do not include molecule formation in our model, we now attempt to divide cold atomic gas from molecular gas by setting a density threshold of $100 \cucm$, above which the timescale for molecule formation is $\lesssim 10 \Myr$ \citep{hws71}. Of this gas we identify as ``molecular'', $51 \%$ has cooled to the minimum temperature ($10 \K$). Cold gas at $n < 100 \cucm$, which we identify as ``CNM'', is shown with the dashed blue line in Figure~\ref{fig:pressure_temp}. The volume-weighted distribution of cold, $n < 100 \cucm$ gas is similar in width to the \citet{jt11} observations but the most probable value is $\approx 0.5 \textrm{ dex}$ lower; our {\em molecular} ($n > 100 \cucm$) gas has a similar median thermal pressure to that observed in the {\em atomic} gas. However, we note that such dense gas is, in reality, often part of larger, self-gravitating structures that will increase its pressure, physics that is not included in our models.

We also note that, by construction, our ``molecular'' gas must have a thermal pressure $\pth/k > 1000 \cucm \K$, whereas our ``CNM'' gas at the maximum temperature commonly found of $\approx 30 \K$ (see histograms in Fig.~\ref{fig:hists_dm}) has a maximum thermal pressure $\pth/k = 3000 \cucm \K$. Therefore, we can learn relatively little by dividing molecular from atomic gas by density in our simulations.

With that strong caveat, we investigate the extent to which star-forming clouds can form in our simulations, absent self-gravity. JM06 argued that the triggered star formation rate in their hydrodynamical simulations was $\lesssim 10 \%$ of the input star formation rate. \citet{kbh09} found that PDFs of column densities in nearby molecular clouds are generally lognormal, suggesting that turbulent motions dominate. However, active star-forming clouds have power-law wings at high column density. \citet{kbb11} argue that the non-lognormal tail represents a transition from diffuse clouds dominated by turbulence to gravitationally bound star-forming clouds. Our models show no evidence for non-lognormality, as expected given the lack of self-gravity. The distribution of the logarithm of densities (which we expect to be Gaussian if the column density distribution is lognormal; \citealt{vg01}) of cold ($T < 200 \K$) gas in our simulations is symmetric about the median density $\langle n \rangle = 43 \cucm$, with a skewness of $\approx -0.04$. This distribution is somewhat flattened compared to a lognormal, with a kurtosis of $\approx -0.4$ (set such that a Gaussian has zero kurtosis). Our molecular gas is part of a continuous distribution with the cold atomic gas. Therefore, we argue that our simulation does not contain the physics necessary for molecular cloud formation: colliding flows in supernova-driven turbulence alone are insufficient to drive significant star formation.

We now consider the relative roles of thermal and turbulent pressure on small ($\sim 8 \pc$) and large ($\sim 125 \pc$) size scales. These pressures are presented in Section~\ref{sec:pressure} and shown in Figure~\ref{fig:phase}$b$ and $c$. On small scales, the turbulent pressure in the warm gas is $\approx 1 \textrm{ dex}$ below the thermal pressure in the warm gas, while the large scale turbulent pressure is comparable to the thermal pressure in the warm gas, with median values of $\log ( \pth/k \, [\textrm{cm}^{-3} \K]^{-1}) = 3.5$ and $\log ( \Pturb{125}/k \, [\textrm{cm}^{-3} \K]^{-1}) = 3.4$. This result is consistent with the claim of \citet{wmh03} that thermal pressure in the WNM is larger than turbulent pressure on scales smaller than $\approx 215 \pc$. They expect the time between supernova-induced shocks in the plane to be greater than the cooling time, so the CNM and WNM should be in approximate thermal pressure equilibrium even in the presence of supernova-driven turbulence.

\subsection{Vertical stratification} \label{sec:stratification}

Our models produce a stratified medium. The medium is in approximate hydrostatic equilibrium: the {\em total} pressure including thermal, magnetic, magnetic tension, and kinetic components of each layer balances the weight of the gas above.

The cold gas is confined close to the plane, with the warmer phases distributed with increasing scale heights. The effects of the magnetic field on vertical stratification are most evident in the cold gas, the only temperature regime in which magnetic pressure dominates over thermal pressure (Section~\ref{sec:pressure} and Fig.~\ref{fig:phase}). The scale height of the cold gas in the \bxf\ model (solid blue line in Fig.~\ref{fig:scale_height}) is $\sim 20 \%$ higher than that of the cold gas in the \bxz\ model (dashed blue line), particularly over the first $\sim 200 \Myr$ of the simulation. At late times, the field in the \bxf\ model has partially dissipated away (see Section~\ref{sec:dynamo}), and the difference between the cold gas scale heights reduces. This can be seen by comparing Figures~\ref{fig:scale_height} and \ref{fig:benergy}. In all of the warmer temperature regimes, the scale heights of each component are comparable in the magnetized (\bxf\ and \bxo) and unmagnetized (\bxz) runs. The negligible role of the field in providing vertical support in the gas warmer than $200 \K$ is consistent with the phase diagrams in Figure~\ref{fig:phase}: the magnetic pressure is the highest fraction of the total pressure in the cold, dense gas.

The scale heights of the two dominant (by mass) phases of the simulated ISM are closely related to the scale heights of the physical processes we imposed upon the simulation (summarized in Table~\ref{tbl:experimental_setup}). The cold gas scale height ($8 \pc$ in the unmagnetized model and $10-11 \pc$ in the magnetized models) is comparable to the characteristic scale height of the gravitational acceleration near the plane ($z \ll z_0$), where $a_1 + a_2 = 8.77 \pc^{-1} \, (\mathrm{pc} \Myr^{-2})^{-1}$. We have written the units here to emphasize that the $a_1$ and $a_2$ parameters in equation~(\ref{eq:pot}) are the linear coefficients describing the change in gravitational acceleration (units of $\mathrm{pc} \Myr^{-2}$) with height (units of \pc). The warm gas scale height ($190-200 \pc$ in all models) is comparable to the core-collapse supernova ($90 \pc$) and diffuse heating ($300 \pc$) scale heights. Because the warm gas is thermally stable primarily due to the balance of radiative cooling and diffuse heating (the grey line in Figure~\ref{fig:phase}$b$), the warm gas scale height is likely tied more strongly to the diffuse heating than to the supernovae.

In our models, the vertical support is provided primarily by turbulent and thermal pressure, with magnetic pressure playing only a small role in even the magnetized runs (Section~\ref{sec:vertical_support} and Figures~\ref{fig:hydrostatics_bx50} and \ref{fig:hydrostatics_bx0}). This contrasts with models that include rotational shear with \citep{gez08} or without \citep{po07} supernovae; in these models, the magnetic pressure does provide significant vertical support. Without supernovae driving turbulence, the role of turbulent pressure is small \citep{po07}.

\subsubsection{Transition-temperature gas}

The near-unity volume and mass filling fractions of transition-temperature gas at $|z| \approx 1 \kpc$ (dashed lines in Figs.~\ref{fig:profile_by_run} and \ref{fig:fillfrac_height}) deserve special attention. Because of the high cooling efficiency of \ion{O}{6} and other high ions, gas at $T \sim 10^5 \K$ is normally assumed to be in transition, either cooling or heating, at an interface between hot and warm layers. The images of density and temperature at $|z| = 1-2 \kpc$ (Fig.~\ref{fig:images}) appear strongly mixed, with slightly less mixing in the magnetized cases than in the unmagnetized case. We also note the distributions of temperature within the $10^{4.5}-10^{5.5} \K$ regime at $|z| \sim 500 \pc$ (green lines in Fig.~\ref{fig:hists_dm}). In the unmagnetized model, the most probable temperature is $\approx 10^{4.8} \K$, a local minimum in our cooling curve (Fig.~1 of JM06), whereas the most probable temperature is $\approx 10^{5.5} \K$ in the magnetized case. This is consistent with previous numerical work that found large amounts of \ion{O}{6}-absorbing gas in supernova cavities in the plane \citep{ab05a}; they argued that turbulent mixing produces a large number of interfaces over a large volume. The suppression of turbulent mixing by the field thus reduces the efficiency of cooling in a turbulent plasma.

In this interpretation, the transition-temperature gas is shock-heated and cooling. Our simulations do not include Lagrangian tracer particles, so we cannot identify the past history of transition-temperature gas. However, because our diffuse heating is turned off at temperatures $T > 2 \times 10^4 \K$, the only heating mechanism capable of producing $10^5 \K$ gas is supernovae and the shocks they produce, suggesting that the widespread transition-temperature gas at $|z| = 1-2 \kpc$ is most likely heated by repeated shocks propagating upwards from the plane.

The observed scale heights of ions that trace transition-temperature gas, determined by comparing the vertical component of the column densities of these ions to the height above the plane of the stars towards which the column densities are measures, are $\sim 3 \kpc$ \citep{sw09}. This is more than three times the $0.6-1 \kpc$ scale height in our models, found by directly measuring the vertical distribution of the transition-temperature gas. A direct comparison of synthetic and real observations will provide the best test of whether this apparent disagreement is real but is beyond the scope of this paper.

\subsubsection{Oscillation} \label{sec:oscillation}

In all cases, the scale heights oscillate. This behavior was predicted in one-dimensional models by \citet{wc01}. In their models, compressions near the midplane drive oscillations in the vertical extent of the ISM. Material near the plane has a ``bounce time'' of $40-60 \Myr$, while the bounce time for higher material is $\sim 100 \Myr$. This behavior can be seen in the images of the evolving halo (Fig.~\ref{fig:images_evol}), in which gas within a few kpc of the plane rises initially and returns to the disk after $60-90 \Myr$, whereas the hot shocks which extend to $\sim 10 \kpc$ return to the disk after $\sim 100-200 \Myr$. Comparison of the scale heights in Figure~\ref{fig:scale_height} shows similar behavior: an initial bounce of the hot gas (red lines) returns to a minimum (the most-compressed mode of \citeauthor{wc01}) after $120 \Myr$, while the warm gas (black lines) oscillates with a shorter period. We reiterate that our star formation rate is constant, so no additional star formation is triggered in the compressed mode. In addition, the amplitude of the oscillations is damped in the magnetized case. This is particularly obvious in the cold gas, which is also the most magnetically dominated.

The \citet{wc01} model also provides a framework for understanding the delayed second set of shocks we discussed in Section~\ref{sec:time_evol}. After the end of the first oscillation of the gas close to the plane at $t \approx 100 \Myr$, a significant column of gas continues to fall from above, leading to a stationary shock which we see at $|z| \approx 3 \kpc$. Once a sufficient column density has accumulated near this shock, a second oscillation is triggered. This occurs at $t \approx 150 \Myr$ in our models.

\subsection{Energy and generation of the magnetic field} \label{sec:dynamo}

The total energy (kinetic, thermal, and magnetic) contained within our box is constant to within $\sim 20 \%$ after $20 \Myr$, suggesting that the turbulence is well developed. We now focus our discussion on the \bxzhr\ and \bxfhr\ models because the \bxo\ model is run at low resolution and its field had not reached a statistical steady-state by the time the run finished after $260 \Myr$. The decrease in magnetic energy over time (Fig.~\ref{fig:benergy}) is likely due to numerical dissipation, with a floor in the magnetic energy established by a turbulent dynamo. The kinetic and thermal energies in the \bxz\ and \bxf\ are comparable throughout the runs (as shown for the thermal energy in Fig.~\ref{fig:benergy}). The total energy in the \bxf\ model is larger than that in the unmagnetized run, with the difference in total energy comparable to the initial magnetic energy.

Most measurements indicate that the magnetic field in the Milky Way consists of an ordered, azimuthal field of $1.5-2.0 \uG$ with an rms field of $5-6 \uG$ \citep[see review by][]{kz08}. After $340 \Myr$, the mass-weighted rms magnetic field strength in the \bxfhr\ model is $4 \uG$, compared to the initial value of $\vec{B}_\mathrm{ini} = (6.5, 0, 0) \uG$. The field remains partially aligned with the initial field: the mass-weighted mean magnetic field vector after $340 \Myr$ is $\vec{B} = (2.0, -0.2, -0.2) \uG$, with standard deviations of $(3, 4, 2) \uG$.

The physical mechanism responsible for maintaining the interstellar field in the Galaxy is still debated. In supernova-driven simulations in a fully periodic box without gravity, \citet{bkm04} found that the magnetic field amplifies with an $e$-folding time of $\sim 10 \Myr$ once turbulence is fully developed. Their simulations did not run for sufficient time to reach a steady-state magnetic energy density. To test similar growth, we have run an additional model at $4 \pc$ resolution, \bxt, noting that \citet{ssb10} and \citet{fss11}  have demonstrated effective dynamos with the HLL3R Riemann solver used in FLASH in this work. This model began with magnetic energy $4 \%$ of the thermal energy, similar to the final state of the \citet{bkm04} work. The magnetic and thermal energy densities of the \bxt\ model are shown as a function of time in Figure~\ref{fig:benergy} (dashed and solid red lines, respectively). Unlike the models with stronger initial fields, the field in the \bxt\ model grows initially (after turbulence is established). The field reaches a maximum surface magnetic energy density of $\approx 0.3 \erg \pc^{-2}$ after $40 \Myr$; the total magnetic energy is approximately $18 \%$ of the thermal energy in the box at this point. After $170 \Myr$, the magnetic energy falls, leaving the magnetic energy $8 \%$ of the thermal energy. The reason for this behavior is unclear.

The decrease of the magnetic energy in the \bxo\ and \bxf\ models, which each have initial magnetic energies comparable to or larger than the thermal energy, combined with the increase (compared to the initial condition) of the magnetic energy in the \bxt\ case suggests that the field tends toward a preferred fraction of equipartition. However, the field energy after $\sim 300 \Myr$ of evolution is dependent upon the initial field energy at $4 \pc$ resolution. Noting that \citet{bkm04} used a resolution of $\approx 1 \pc$ in their supernova-driven dynamo model, a high resolution version of the \bxt\ model may prove illuminating.

Given the decrease in the magnetic energy to $\lesssim 1/3$ of equipartition in our models, it is clear that a dynamo driven by supernovae alone will not create a magnetic field with an important role in the vertical stratification of the ISM. Rotational shear can amplify the field exponentially to at least $0.5 \uG$ \citep[Fig. 3 of][]{gez08}, though no interstellar dynamo model to date has been run for long enough to allow the field to reach equipartition. In their models without supernovae, \citet{po07} found that a dynamo driven by differential rotation allows the magnetic energy density to reach $0.3 - 0.6$ of equipartion with the thermal energy density, but the turbulent energy density is quite small absent supernovae.

\subsection{Numerical resolution} \label{sec:numerical_res}

Our approach of running the simulations at medium ($4 \pc$) resolution for $300 \Myr$ before switching to high ($2 \pc$) resolution suggests that our main results are well converged. The mass fractions (Fig.~\ref{fig:smd}) and vertical distributions (Fig.~\ref{fig:scale_height}) in each component of the ISM are modified by $< 1 \%$ between the medium and high resolution runs, indicating that the overall distribution of material, particularly as a function of height, is well converged at the $4 \pc$ midplane resolution. The volume filling fractions in the plane (Fig.~\ref{fig:time_evol}) are affected more significantly, with the volume occupied by cold and warm gas each decreasing by $\approx 10 \%$ in the $5 \Myr$ immediately following the resolution switch. Both volume filling fractions then reach a steady-state at the new resolution. Cold clouds exist at sizes down to the resolution limit at both medium and high resolutions, indicating that we do not resolve the smallest clouds.  However, such collapse may not be physically meaningful without including self-gravity and modeling of the formation of molecules.

\subsection{Missing large-scale effects} \label{sec:missing_largescale}

We model an approximation of physical parameters appropriate for the solar circle. With these parameters, the simulations reach a statistical steady-state only after $\gtrsim 200 \Myr$ in the plane, with oscillations in the halo continuing for the entire run of our simulations. However, several factors related to Galactic structure on scales larger than $1 \kpc$ are missing from our work.

{\em Spiral arms:}
The supernova rate adopted here corresponds roughly to a star formation rate averaged over the Galactic disk. In reality, massive stars form in spiral arms which pass a given parcel of the Galaxy every $\sim 200 \Myr$. Therefore, a more complete model would vary the Type~II supernova rate.

{\em Accretion:}
Our model does not include accretion. High velocity clouds (HVCs), which have typical masses of $\sim 10^6 - 10^7 \, M_\mathrm{\odot}$, provide a mass influx of $\sim 1 M_\mathrm{\odot} \yr^{-1}$ over the entire Galactic disk \citep{wyh07,tpp08}. The interaction of HVCs with the Galactic halo likely has a significant effect on the behavior of the ISM and may drive turbulence. However, the mechanism by which infalling material joins the Galactic ISM is unknown and is beyond the scope of this paper.

{\em Differential rotation:} 
Differential rotation may be an important driver of turbulence, particularly in the outer galaxy through the magnetorotational instability \citep{po07}. Its inclusion may allow some form of large scale dynamo to function, as discussed in Section~\ref{sec:dynamo}.

\section{Conclusions} \label{sec:sn_mhd_conclusions}

We have considered the role of supernovae in the vertical stratification of components of a magnetized ISM in collisional ionization equilibrium in the Galactic gravitational field using stable, positivity-preserving 3D AMR MHD simulations. We summarize our main conclusions here.

\begin{enumerate}
\item Our models contain distinct cold and warm phases of the ISM (Figs.~\ref{fig:hists_dm}, \ref{fig:hists_dv}, and \ref{fig:phase}$b$). Reasonable treatment of the diffuse heating rate, which models photoelectric heating of gas cooler than $2 \times 10^4 \K$, is necessary for distinct phases to be present. With excessive diffuse heating, stable equilibria do not exist (JM06).
\item The qualitative structure of the ISM is similar in magnetized and unmagnetized models (Fig.~\ref{fig:images}). The field in our models also has very little effect on the vertical distribution or scale heights of the components of the ISM (Fig.~\ref{fig:profile_by_phase}). This is because the horizontally-averaged magnetic pressure ($|\vec{B}|^2 /2$) and magnetic tension force ($-B_z^2$) nearly cancel each other out at all heights (Fig.~\ref{fig:hydrostatics_bx50}$a$). Even ignoring magnetic tension, the magnetic pressure dominates over the thermal pressure only in the cold gas, and the cold gas is the only phase which shows some evidence for an increased scale height in the magnetized model. This effect is small because kinetic pressure is dominant in the cold gas in our models (Fig.~\ref{fig:phase}$d$ and $a$).
\item Hydrostatic equilibrium is maintained primarily by thermal and kinetic pressure. We do not confirm the hypothesis of \citet{whj10} that magnetic fields would help to address the smaller-than-observed scale height of the warm ionized medium in the hydrodynamical models used in that work. In both cases, we do not model photoionization heating, which could lead to additional vertical support for the WIM. Rotational shear may well be important in allowing the magnetic field to provide vertical support \citep{po07,gze08}.
\item The addition of a field reduces the mass in cold gas by $8 \%$ with a corresponding increase in the mass contained in lower density, warm gas (Fig.~\ref{fig:smd}), consistent with our expectation that the magnetic field would reduce the compressibility of the gas. The volume filling fraction of the cold gas in the plane is also $\approx 20 \%$ higher in the $2 \pc$ resolution, magnetized model than in the similar unmagnetized one (thick, blue lines in Fig.~\ref{fig:time_evol} and blue lines in Fig.~\ref{fig:fillfrac_height}).
\item Extending the box to heights $|z| \gtrsim 10 \kpc$ to fully capture the Galactic fountain flow is the single most important numerical factor in establishing a realistic temperature distribution in the halo (Section~\ref{sec:box_size}).
\item Even though our star formation rate does not change over time, the gas oscillates vertically (Figs.~\ref{fig:images_evol} and \ref{fig:scale_height}), as predicted by \citet{wc01}.
\item At heights $1 \kpc \lesssim |z| \lesssim 2 \kpc$, the modelled ISM consists primarily of thermally unstable ``transition-temperature'' ($T \sim 10^5 \K$) gas (Fig.~\ref{fig:profile_by_run}). This is a thick transition region between the disk---mostly warm gas with cold clouds and hot supernova remnants---and the hot halo. In this transition region, the gas is strongly mixed, with interfaces between hot and warm gas filling most of the volume, leading to the near-unity filling fraction of $10^5 \K$ gas.
\end{enumerate}

\acknowledgements

The authors thank  K.\ Wood for helpful discussions and E.\ Jenkins for a comment on the \ion{C}{1} distribution. A.~S.~H.\ thanks S.~Friedman and B.~Morsony for sharing their expertise with FLASH, his thesis committee (M.~A.~Bershady, J.~S.~Gallagher, D.~McCammon, and E.~M.~Wilcots) for comments on a draft of this paper, and the American Museum of Natural History for hospitality during the initial stages of this work. The FLASH code was developed by the DOE-supported ASC/Alliance Center for Astrophysical Thermonuclear Flashes at the University of Chicago. The computations were performed using National Science Foundation-supported TeraGrid resources provided by the Texas Advanced Computing Center under grant number TG-MCA99S024. This work was partly supported by NASA/SAO grant TM0-11008X, NASA ATP grant NNX10A170G, and NSF grants AST-1109395 and AST-0607512.

\bibliography{references}

\end{document}